\documentclass{svjour2}                   
\smartqed 
\usepackage{graphicx}

\usepackage{epsfig}

\journalname{Foundations~of~Physics}

\begin{document}

\title{Making the Case for Conformal Gravity}

\author{Philip D. Mannheim}

\institute{Department of Physics,
University of Connecticut, Storrs, CT 06269, USA \\
\email{philip.mannheim@uconn.edu} }

\date{October 27, 2011}

\maketitle

\begin{abstract}

We review some recent developments in the conformal gravity theory that has been advanced as a candidate alternative to standard Einstein gravity.  As a quantum theory the conformal theory is both renormalizable and unitary, with unitarity being obtained because the theory is a $PT$ symmetric rather than a Hermitian theory. We show that in the theory there can be no a priori classical curvature, with all curvature having to result from quantization. In the conformal theory gravity requires no independent quantization of its own, with it being quantized solely by virtue of its being coupled to a quantized matter source. Moreover, because it is this very  coupling that fixes the strength of the gravitational field commutators, the gravity sector zero-point energy density and pressure fluctuations are then able to identically cancel the zero-point fluctuations associated with the matter sector. In addition, we show that when the conformal symmetry is spontaneously broken, the zero-point structure automatically readjusts so as to identically cancel the cosmological constant term that dynamical mass generation induces. We show that the macroscopic classical theory that results from the quantum conformal theory incorporates global physics effects that provide for a detailed accounting of a comprehensive set of 138 galactic rotation curves with no adjustable parameters other than the galactic mass to light ratios, and with the need for no dark matter whatsoever. With these global effects eliminating the need for dark matter, we see that invoking dark matter in galaxies could potentially be nothing more than an attempt to describe global physics effects in purely local galactic terms. Finally, we review some recent work by 't Hooft in which a connection between conformal gravity and Einstein gravity has been found.


\end{abstract}

\section{Einstein gravity: what must be kept}
\label{S1}

Following his development of special relativity, Einstein was faced with two immediate problems. The first was to make Newtonian gravity compatible with relativity, and the second was to develop a formalism in which not only uniformly moving observers but also accelerating ones would all be able to agree on the same physics. While these two issues are logically independent (even in the absence of gravity one has to able to write Newton's second law of motion in an observer-independent way), by imposing general coordinate invariance and by identifying the spacetime metric $g_{\mu\nu}(x)$ as the gravitational field, Einstein was able to provide a solution to both problems simultaneously. In this formalism a central role is played by the Christoffel symbols
\begin{equation}
\Gamma^{\mu}_{\nu\sigma}={1 \over 2}g^{\mu\lambda}\left[\partial_{\nu}g_{\lambda\sigma} +
\partial_{\sigma}g_{\lambda\nu} -\partial_{\lambda}g_{\nu\sigma}\right],
\label{M!}
\end{equation}
since in terms of them one can show that the path that minimizes the distance 
\begin{equation}
ds^2=g_{\mu\nu}(x)dx^{\mu}dx^{\nu}
\label{M2}
\end{equation}
between two points in any geometry (curved or flat) is the one that obeys the geodesic equation
\begin{equation}
{d^2x^{\mu} \over ds^2}+\Gamma^{\mu}_{\nu\sigma}{dx^{\nu}\over ds}{dx^{\sigma} \over ds}=0.
\label{M3}
\end{equation}
The key feature of this equation is that even though neither of the two terms that appears in (\ref{M3}) is itself a true coordinate vector, the linear combination of them is, with the vanishing of their sum in any given coordinate system ensuring its vanishing in any other. 

Even though the Christoffel symbols  are not themselves true coordinate tensors, from them one can construct a quantity that is, viz. the Riemann tensor as defined by
\begin{equation}
R^{\lambda}_{\phantom{\lambda}\mu\nu\kappa}={\partial\Gamma^{\lambda}_{\mu\nu} \over \partial x^{\kappa}}+\Gamma^{\lambda}_{\kappa\eta}\Gamma^{\eta}_{\mu\nu}
-{\partial\Gamma^{\lambda}_{\mu\kappa} \over \partial x^{\nu}}-\Gamma^{\lambda}_{\nu\eta}\Gamma^{\eta}_{\mu\kappa}.
\label{M4}
\end{equation}
The utility of this tensor is that a given spacetime will be flat if and only if every component of $R^{\lambda}_{\phantom{\lambda}\mu\nu\kappa}$ is zero. When all components of $R^{\lambda}_{\phantom{\lambda}\mu\nu\kappa}$ are zero, (\ref{M3}) describes Newton's second law of motion for a free particle in the absence of gravity as viewed in an accelerating coordinate system. And when $R^{\lambda}_{\phantom{\lambda}\mu\nu\kappa}$ is non-zero, there is a choice of values for the Christoffel symbols (viz. the Schwarzschild metric ones that are associated with the vanishing of the Ricci tensor $R_{\mu\kappa}=R^{\lambda}_{\phantom{\lambda}\mu\lambda\kappa}$) that enables (\ref{M3}) to describe Newton's law of gravity, again in an arbitrary accelerating coordinate system. Then, with the Schwarzschild metric also giving relativistic corrections to Newtonian gravity, the observation of the predicted gravitational bending of light by the Sun established the validity of the above description of nature.

One can thus say with confidence that gravity is a covariant metric theory in which the metric  describes the gravitational field, and that the geometry in the vicinity of the Sun is given by the Schwarzschild metric 
\begin{equation}
ds^2=-B(r)dt^2+A(r)dr^2+r^2d\theta^2+r^2\sin^2\theta d\phi^2,
\label{M5}
\end{equation}
where
\begin{equation}
B(r)=A^{-1}(r)=1-2\beta/r,~~~~~\beta=MG/c^2,
\label{M6}
\end{equation}
up to the perturbative order  to which the metric has so far actually been tested. Thus any viable theory of gravity must embody all of the above, and we note that in the above we have not specified the equation of motion that is to be obeyed by the gravitational field. Rather, we have indicated only what its solution on solar system distance scales needs to look like.

\section{Einstein gravity: what could be changed}
\label{S2}

To complete the theory one thus needs to specify the gravitational field equations themselves. To this end Einstein postulated that  the needed equations are to be of the form
\begin{equation}
-\frac{1}{8\pi G}\left(R^{\mu\nu} -\frac{1}{2}g^{\mu\nu}R^{\alpha}_{\phantom{\alpha}\alpha}\right)=T^{\mu\nu}_{\rm M},
\label{M7}
\end{equation}
a set of equations that can be obtained via functional variation with respect to the metric of an action for the Universe of the form
\begin{equation}
I_{\rm UNIV}=I_{\rm EH}+I_{\rm M}=-{1 \over 16\pi G}\int d^4x (-g)^{1/2}R^{\alpha}_{\phantom{\alpha}\alpha}+I_{\rm M}.
\label{M8}
\end{equation}
In (\ref{M7}) and (\ref{M8})  ${\rm M}$ denotes the matter field sector and $T^{\mu\nu}_{\rm M}$ is the matter field energy-momentum tensor. Given (\ref{M7}), we immediately see that in the source-free region where $T^{\mu\nu}_{\rm M}=0$ the Ricci tensor has to vanish, with the solution given in (\ref{M5}) and (\ref{M6}) then following in the region exterior to a static, spherically symmetric source.

However, while (\ref{M5}) and (\ref{M6}) follow from (\ref{M7}), this is not the only way to secure the Ricci-flat Schwarzschild solution. Consider for example an action for the Universe of the conformal form
\begin{eqnarray}
I_{\rm UNIV}=I_{\rm W}+I_{\rm M}&=&
-\alpha_g\int d^4x (-g)^{1/2}C_{\lambda\mu\nu\kappa} C^{\lambda\mu\nu\kappa}+I_{\rm M}
\nonumber\\
&\equiv&-2\alpha_g\int d^4x (-g)^{1/2}\left[R^{\mu\nu}R_{\mu\nu}-{1 \over 3}(R^{\alpha}_{\phantom{\alpha}\alpha})^2\right] +I_{\rm M},
\label{M9}
\end{eqnarray}
where
\begin{eqnarray}
C_{\lambda\mu\nu\kappa}&=&R_{\lambda\mu\nu\kappa}
+{1 \over 6}R^{\alpha}_{\phantom{\alpha}\alpha}\left[
g_{\lambda\nu}g_{\mu\kappa} -g_{\lambda\kappa}g_{\mu\nu}\right]
\nonumber\\
&-&{1 \over 2}\left[
g_{\lambda\nu}R_{\mu\kappa} -g_{\lambda\kappa}R_{\mu\nu}
-g_{\mu\nu}R_{\lambda\kappa} +g_{\mu\kappa}R_{\lambda\nu}\right]
\label{M10}
\end{eqnarray}
is the Weyl conformal tensor. Functional variation of this action with respect to the metric leads to the equation of motion (see e.g. \cite{Mannheim2006})
\begin{equation}
-4\alpha_g W^{\mu\nu}+T^{\mu\nu}_{\rm M}=0,
\label{M11}
\end{equation}
where
\begin{eqnarray}
&&W^{\mu \nu}=
\frac{1}{2}g^{\mu\nu}(R^{\alpha}_{\phantom{\alpha}\alpha})   
^{;\beta}_{\phantom{;\beta};\beta}+
R^{\mu\nu;\beta}_{\phantom{\mu\nu;\beta};\beta}                     
 -R^{\mu\beta;\nu}_{\phantom{\mu\beta;\nu};\beta}                        
-R^{\nu \beta;\mu}_{\phantom{\nu \beta;\mu};\beta}                          
 - 2R^{\mu\beta}R^{\nu}_{\phantom{\nu}\beta}                                 
\nonumber \\
&&+\frac{1}{2}g^{\mu\nu}R_{\alpha\beta}R^{\alpha\beta}-
\frac{2}{3}g^{\mu\nu}(R^{\alpha}_{\phantom{\alpha}\alpha})          
^{;\beta}_{\phantom{;\beta};\beta}                                              
+\frac{2}{3}(R^{\alpha}_{\phantom{\alpha}\alpha})^{;\mu;\nu}                           
+\frac{2}{3} R^{\alpha}_{\phantom{\alpha}\alpha}
R^{\mu\nu}                              
-\frac{1}{6}g^{\mu\nu}(R^{\alpha}_{\phantom{\alpha}\alpha})^2,
\nonumber \\
\label{M12}
\end{eqnarray}        
to thus yield a gravitational theory that also has $R_{\mu\nu}=0$ as a vacuum solution.\footnote{Since $W^{\mu\nu}$ is constructed as the functional variation with respect to the metric of a gravitational action that is both general coordinate invariant and locally conformal invariant, it kinematically obeys $W^{\mu\nu}_{\phantom{\mu\nu};\nu}=0$, $g_{\mu\nu}W^{\mu\nu}=0$. Likewise a matter energy momentum tensor constructed as the functional variation with respect to the metric of a matter action  that is general coordinate and locally conformal invariant will obey $T_{\rm M~;\nu}^{\mu\nu}=0$, $g_{\mu\nu}T_{\rm M}^{\mu\nu}=0$. While the tracelessness of $T_{\rm M}^{\mu\nu}$ forbids the matter fields from having kinematical or  mechanical masses, it is important to  note \cite{Mannheim2006} that it it does not prevent them from acquiring dynamical masses via spontaneous symmetry breaking. Specifically, as for instance seen explicitly in Sec. (7) below, the scalar order parameter $\langle S|\bar{\psi}\psi|S\rangle$ that gives a fermionic matter field a mass also carries energy density and momentum that serve to maintain  $g_{\mu\nu}T_{\rm M}^{\mu\nu}=0$. Moreover, since in flat space it does this by adding a cosmological constant term $T_{\rm COS}^{\mu\nu}$ on to the standard kinematic perfect fluid $T_{\rm KIN}^{\mu\nu}$ (see e.g. (\ref{M45}) below and \cite{Mannheim2006}), it does not affect energy differences or the conservation of $T_{\rm KIN}^{\mu\nu}$. In flat space $T_{\rm COS}^{\mu\nu}$ is not observable, and one can use the non-traceless $T_{\rm KIN}^{\mu\nu}$ to describe macroscopic systems. It is only in its coupling to gravity that presence of $T_{\rm COS}^{\mu\nu}$ can be felt. } This analysis thus shows that the Einstein equations given in (\ref{M7}) are only sufficient to give the Schwarzschild solution and its non-relativistic Newtonian limit but not necessary. 

The Einstein equations are thus not uniquely selected. Moreover, in and of itself, the requirement that the gravitational action be a general coordinate scalar does not at all restrict the gravitational sector of the action to be of the Einstein-Hilbert $I_{\rm EH}$ form given in (\ref{M8}), with the number of possible general coordinate invariant gravitational actions that one could write down actually being infinite, since one could use arbitrarily high powers of the Riemann tensor and its contractions. This lack of uniqueness is familiarly reflected in the fact that one is free to augment (\ref{M8}) with a term of the form $-\int d^4x (-g)^{1/2}\Lambda$ where $\Lambda$ is the cosmological constant. 

Beyond this we note that if one takes (\ref{M7}) as a given and extrapolates it beyond its weak classical gravity solar system origins, one runs into difficulties in essentially every type of extrapolation that one could consider. Thus if one extrapolates the classical theory to galaxies and clusters of galaxies one runs into the dark matter problem, if one extrapolates the classical theory to cosmology one runs into the cosmological constant or dark energy problem, if one extrapolates to strong classical gravity one runs into the singularity problem, and if one quantizes the theory and extrapolates far off the mass shell one runs into the renormalizability and zero-point problems. Now even though no dark matter has yet been detected and dark energy is not yet at all understood, if one nonetheless takes dark matter and dark energy as a given, one then encounters many successes as well (such as big bang nucleosynthesis, anisotropy of the cosmic microwave background, strong lensing). However, to achieve these successes one has to take the energy budget of the Universe to be of order 70\% dark energy  and 25\% dark matter, with only 5\% or so being regular luminous baryonic matter. Not only is there as yet no detection of the needed dark matter particles, the required amount of dark energy is 60 orders of magnitude or so less than the amount expected from fundamental elementary particle physics -- and if one were to use the large particle physics value the fits would be disastrous. Moreover, with all applications to date of gravity to astrophysics and cosmology having been made with gravity itself being treated classically, there is no guarantee that the current successes of standard Einstein gravity would not be modified by its non-renormalizable quantum corrections. Given these concerns we shall thus look for a completely different extrapolation of solar system wisdom, in a theory that is unambiguously specified. As we shall see, via the imposition of a particular invariance principle, namely local conformal invariance, none of the above extrapolation problems or ambiguities will any longer be encountered.

\section{Conformal gravity: an ab initio approach}
\label{S3}

To see how things work, we note that if we start with the kinetic energy of a free massless fermion in flat spacetime, then in order to obtain an action that is locally gauge invariant under $\psi(x)\rightarrow e^{i\beta(x)}\psi(x)$ (we suppress internal symmetry group indices), we introduce a gauge field that transforms as $A_{\mu}(x)\rightarrow A_{\mu}(x)+\partial_{\mu}\beta(x)$, and minimally couple according to 
\begin{equation}
I_{\rm M}=-\int d^4x\bar{\psi}(x)\gamma^{\mu}[i\partial_{\mu}+A_{\mu}(x)]\psi(x). 
\label{M13}
\end{equation}
Similarly, if we start with the kinetic energy of a free massless fermion in flat spacetime, in order to obtain an action that is locally coordinate invariant, we introduce the fermion spin connection $\Gamma_{\mu}(x)$, with the action taking the form
\begin{equation}
I_{\rm M}=-\int d^4x (-g)^{1/2}\bar{\psi}(x)\gamma^{\mu}(x)[i\partial_{\mu}+i\Gamma_{\mu}(x)]\psi(x),
\label{M14}
\end{equation}
with $\gamma^{\mu}(x)=V^{\mu}_a(x)\hat{\gamma}^a$ and $\Gamma_{\mu}(x)=([\gamma^{\nu}(x),\partial_{\mu}\gamma_{\nu}(x)]-[\gamma^{\nu}(x),\gamma_{\sigma}(x)]\Gamma^{\sigma}_{\mu\nu})/8$, and with $V^{\mu}_a(x)$ being a vierbein and the four $\hat{\gamma}^a$ being the special-relativistic Dirac gamma matrices. 

Having now obtained (\ref{M14}) this way, we find that without having required it explicitly, this action actually has an additional symmetry, as it is locally conformal invariant under $\psi(x)\rightarrow e^{-3\alpha(x)/2}\psi(x)$, $V_{\mu}^a(x)\rightarrow e^{\alpha(x)}V_{\mu}^a(x)$, $g_{\mu\nu}(x)\rightarrow e^{2\alpha(x)}g_{\mu\nu}(x)$. Consequently, we can regard the spin connection $\Gamma_{\mu}(x)$ as being introduced not to maintain local coordinate invariance but rather to maintain local conformal invariance instead, in exactly the same minimally-coupled way that $A_{\mu}(x)$ maintains local gauge invariance. Thus if we require that the kinetic energy of a massless fermion be invariant under complex phase transformations of the form $\psi(x)\rightarrow e^{-3\alpha(x)/2+i\beta(x)}\psi (x)$, we will be led to an action
\begin{equation}
I_{\rm M}=-\int d^4x (-g)^{1/2}\bar{\psi}(x)\gamma^{\mu}(x)[i\partial_{\mu}+i\Gamma_{\mu}(x)+A_{\mu}(x)]\psi(x)
\label{M15}
\end{equation}
that is locally gauge invariant and locally conformal invariant combined. (Under a local gauge transformation the metric transforms as $g_{\mu\nu}(x)\rightarrow g_{\mu\nu}(x)$, while under a local conformal transformation the gauge field transforms as $A_{\mu}(x)\rightarrow A_{\mu}(x)$.)

The reason why such a local conformal structure emerges in (\ref{M14}) is that massless particles move on the light cone, and the light cone is not just Poincare invariant, it is invariant under the full 15-parameter conformal group $SO(4,2)$. (If the  $ds^2=g_{\mu\nu}(x)dx^{\mu}dx^{\nu}$ line element is zero, then so is $ds^2=e^{2\alpha(x)}g_{\mu\nu}(x)dx^{\mu}dx^{\nu}$.) Moreover, the covering group of $SO(4,2)$ is $SU(2,2)$. Since $SU(2,2)$ is generated by the 15 Dirac matrices ($\gamma_5$, $\gamma_{\mu}$, $\gamma_{\mu}\gamma_5$, $[\gamma_{\mu},\gamma_{\nu}]$), its fundamental representation is a fermionic field, and the full conformal structure of the light cone is thus built into a massless fermionic field, with a 4-component Dirac spinor being irreducible under the conformal group even as it is reducible under Poincare.

As we see, it is thus natural to take fermions to be the most basic elements in physics, with internal symmetry gauge fields and a gravitational spin connection being induced once one gives the fermion kinetic energy a local complex phase invariance. However, such a starting point does not generate any kinetic energy terms for the gauge and gravitational fields. To actually generate them rather than just postulate them we take note of a calculation by 't Hooft \cite{'tHooft2010}. Specifically, 't Hooft  evaluated the logarithmically divergent part of the path integral $\int D\bar{\psi}D\psi\exp(iI_{\rm M})$ associated with (\ref{M15}), and found that after dimensional regularization it took the form of an effective action:
\begin{equation}
I_{\rm EFF}=\int d^4x (-g)^{1/2}C\left[\frac{1}{20}[R^{\mu\nu}R_{\mu\nu}-{1 \over 3}(R^{\alpha}_{\phantom{\alpha}\alpha})^2]
+\frac{1}{3}F_{\mu\nu}F^{\mu\nu}\right],
\label{M16}
\end{equation}
where $C=1/8\pi^2(4-D)$ in spacetime dimension $D$. Comparing with (\ref{M9}) we see that we have generated none other than the conformal gravity action $\int d^4x (-g)^{1/2}C_{\lambda\mu\nu\kappa} C^{\lambda\mu\nu\kappa}$ (as then rewritten using the Gauss-Bonnet theorem) together with the Maxwell action, and take note of the fact that this procedure did not generate the Einstein-Hilbert action. With the Maxwell action being invariant under $A_{\mu}(x)\rightarrow A_{\mu}(x)+\partial_{\mu}\beta(x)$ and with the conformal gravity action being invariant under $g_{\mu\nu}(x)\rightarrow e^{2\alpha(x)}g_{\mu\nu}(x)$ (see e.g. \cite{Mannheim2006}), the conformal gravity action thus serves as the gravitational analog of the Maxwell action, and in the following we shall thus use local conformal invariance as the principle with which to fix the structure of the gravitational action. In so doing we see that gravity can be generated by gauging the full conformal symmetry of the light cone.

Given the assumption of local conformal invariance, we find that the conformal gravity action $I_{\rm W}=-\alpha_g\int d^4x (-g)^{1/2} C_{\lambda\mu\nu\kappa} C^{\lambda\mu\nu\kappa}$ is the unique gravitational action that is invariant under the local transformation $g_{\mu\nu}(x)\rightarrow e^{2\alpha(x)}g_{\mu\nu}(x)$, with the gravitational coupling constant $\alpha_g$ being dimensionless. Because the coupling constant $\alpha_g$ is dimensionless, the conformal theory is power-counting renormalizable, and thus if we did not include an initial $\int d^4x (-g)^{1/2}C_{\lambda\mu\nu\kappa} C^{\lambda\mu\nu\kappa}$ term in the action, we would anyway generate one as a renormalization counter-term. Because the variation of the conformal  action leads to fourth-order equations of motion, it had long been thought that the theory would not be unitary. However, as we describe in Sec. (\ref{S5}),  Bender and Mannheim \cite{Bender2008a,Bender2008b} have recently shown that one can find a realization of the theory that is unitary. Consequently, conformal gravity is to be regarded as a bona fide quantum gravitational theory. Moreover, the similarity of the theory to Maxwell theory also carries over to the generation of a macroscopic classical limit starting from a microscopic quantum field theory. Specifically, in just the same way as the classical Maxwell equations emerge from the quantum Maxwell equations as matrix elements of the quantum fields in states containing an indefinite number of photons, because of its renormalizability the same will happen for conformal gravity  in states containing an indefinite number of gravitational quanta, with both the quantum theory and its macroscopic classical limit obeying the equation of motion given in (\ref{M11}). Consequently, in the following we will be able to use (\ref{M11}) to study both the microscopic zero-point fluctuation problem and the macroscopic behavior of the theory on astrophysical distance scales.

The non-renormalizable Einstein-Hilbert action is expressly forbidden by the conformal symmetry because Newton's constant carries an intrinsic dimension. However, as noted above, this does not prevent the theory from possessing the Schwarzschild solution and Newton's law of gravity. In addition, the same conformal symmetry forbids the presence of any intrinsic cosmological constant term as it carries an intrinsic dimension too; with conformal invariance thus providing a very good starting point for tackling the cosmological constant problem.

Now we recall that the fermion and gauge boson sector of the standard $SU(3)\times SU(2)\times U(1)$ model of strong, electromagnetic, and weak interactions is also locally conformal invariant since all the associated coupling constants are dimensionless, and gauge bosons and fermions get masses dynamically via spontaneous symmetry breaking. Other than the Higgs sector (which we shall shortly dispense with), the standard model Lagrangian is devoid of any intrinsic mass or length scales. And with its associated energy-momentum tensor serving as the source of gravity, it is thus quite natural that gravity should be devoid of any intrinsic mass or length scales too. Our use of conformal gravity thus nicely dovetails with the standard $SU(3)\times SU(2)\times U(1)$ model. To tighten the connection, we note that while the standard $SU(3)\times SU(2)\times U(1)$ model is based on second-order equations of motion, an electrodynamics Lagrangian of the form $F^{\mu\nu}\partial_{\alpha}\partial^{\alpha} F_{\mu\nu}$ would be just as gauge and Lorentz invariant as the Maxwell action, and there is no immediate reason to leave any such type of term out. Now while an $F^{\mu\nu}\partial_{\alpha}\partial^{\alpha} F_{\mu\nu}$ theory would not be renormalizable, in and of itself renormalizability is not a law of nature (witness Einstein gravity). However, such a theory would not  be conformal invariant. Thus if we impose local conformal invariance as a principle, we would then force the fundamental gauge theories to be second order, and thus be renormalizable after all. However, imposing the same symmetry on gravity expressly forces it to be fourth order instead, with gravity then also being renormalizable. As we see, renormalizability is thus a consequence of conformal invariance.

Now if the underlying theory is to be locally conformal invariant, there would be no place for a fundamental Higgs field with its tachyonic double-well potential. Instead, mass scales would have to be generated dynamically in the vacuum via dynamical fermion bilinear condensates. The elimination of a fundamental Higgs field has an immediate benefit --  one no longer has to deal with the uncontrollable cosmological constant contribution it produces when it acquires a non-vanishing expectation value. However, one still has to make contact with the standard model, and one would thus want to obtain a standard model Lagrangian with some effective scalar field. Such an effective scalar field would have to emerge as a Ginzburg-Landau c-number order parameter, i.e. as the matrix element of a fermion bilinear operator in some possibly spacetime-dependent coherent state. Such an effective c-number scalar field would not be observable in an accelerator.

To see what such an effective theory might look like, we note that if the fermion acquires  a mass parameter $M(x)$ by some dynamical symmetry breaking mechanism, the associated Hartree-Fock mean-field action would take the same form as given in (\ref{M15}), only with a mass term added, viz.
\begin{equation}
I_{\rm M}=-\int d^4x (-g)^{1/2}\bar{\psi}(x)\gamma^{\mu}(x)[i\partial_{\mu}+i\Gamma_{\mu}(x)+A_{\mu}(x)+M(x)]\psi(x).
\label{M17}
\end{equation}
Evaluating the logarithmically divergent part of the same $\int D\bar{\psi}D\psi\exp(iI_{\rm M})$ path integral as before only with (\ref{M17}) this time  generates the same $I_{\rm EFF}$ as in (\ref{M16}), while adding on the mean-field action \cite{'tHooft2010} (as written here using the sign convention employed in this paper for $R^{\alpha}_{\phantom{\alpha}\alpha}$):
\begin{equation}
I_{\rm MF}=\int d^4x (-g)^{1/2}C\left[-M^4(x)+\frac{1}{6}M^2(x) R^{\alpha}_{\phantom{\alpha}\alpha}-g_{\mu\nu}\partial^{\mu}M(x)\partial^{\nu}M(x)\right].
\label{M18}
\end{equation}
Here $C$ is the same logarithmically divergent constant as before. Finally, if we give $M(x)$ a group index, the same procedure would cause the $\partial^{\mu}M(x)$ terms to be replaced by covariant gauge derivatives (see \cite{Eguchi1974,Mannheim1976}), and would yield
\begin{eqnarray}
I_{\rm MF}&=&\int d^4x (-g)^{1/2}C\bigg{[}-M^4(x)+\frac{1}{6}M^2(x) R^{\alpha}_{\phantom{\alpha}\alpha}
\nonumber\\
&-&g_{\mu\nu}(\partial^{\mu}+iA^{\mu}(x))M(x)(\partial^{\nu}-iA^{\nu}(x))M(x)\bigg{]}.
\label{M19}
\end{eqnarray}
When (\ref{M19}) is taken in conjunction with (\ref{M16}), a conformally coupled standard model emerges, but with a c-number scalar parameter $M(x)$ that is not a fundamental field. In Secs. (\ref{S7}) and (\ref{S8}) we will show how such a mass parameter $M(x)$ could be generated as a fermion bilinear condensate in a conformal invariant theory. And in \cite{Mannheim2009a} we explore the degree to which the existence of such a c-number order parameter $M(x)$ necessitates the existence of an accompanying dynamical bound state scalar particle.

\section{Quantization of gravity through coupling}
\label{S4}

In trying to find solutions to the conformal theory, we note that given the lack of any intrinsic mass or length scales in the conformal action, without dynamical generation of such scales there could be no non-trivial solutions to the theory. Thus if all mass and length scales are to come from dynamics, and all such mass-generating dynamics is to be quantum-mechanical (c.f. no fundamental Higgs fields), the only allowed geometry in purely classical conformal gravity would be one with no curvature at all, viz. a geometry that is Minkowski. (For there to be any curvature one needs some length scale to characterize it.) Thus if one takes $W^{\mu\nu}$ to be a classical fourth-order derivative function, then even though one could readily find an exterior vacuum solution to $W^{\mu\nu}=0$ such as the Ricci-flat one given in (\ref{M5}) and (\ref{M6}), there would be no basis for taking the dimensionful parameter $\beta$ to be non-zero since the theory is as yet scale free. For the solution to a differential equation to have lower symmetry than the equation itself it (i.e. for a solution to contain symmetry-breaking integration constants that are not present in the equation itself), there has to be a spontaneous breakdown of the symmetry. If all symmetry breaking is to be quantum-mechanical, then in the absence of quantum mechanics, geometry would have to be flat. Thus, if the mass that appears in the Schwarzschild radius of a source is quantum-mechanically generated, then the curvature produced by that mass is ipso facto due to quantum mechanics too. 

The above remarks require some clarification since in classical electrodynamics  one can construct plane wave solutions with $k^0=|\bar{k}|$ even though the classical Maxwell equations are conformal invariant. Specifically, if we have a free classical Maxwell action $\int d^4x F_{\mu\nu}F^{\mu\nu}$ in flat spacetime, the associated homogeneous wave equation takes the form $(\partial_t^2-\nabla^2)A_{\mu}(x)=0$ (in the convenient Lorentz gauge). However, while this equation has $A_{\mu}(x)=\epsilon_{\mu}(x)\exp(-ik \cdot x)$ as a solution, this solution involves a dimensionful four-vector momentum $k_{\mu}$ that is not present in the equation of motion itself. Hence some mechanism is required to generate such a four-momentum. In classical Maxwell theory the mechanism for doing this is not from the $F_{\mu\nu}$ sector of the theory at all, but rather via the introduction of a localized source $J_{\mu}(x)$. In the presence of the source  the solution to the inhomogeneous $(\partial_t^2-\nabla^2)A_{\mu}(x)=J_{\mu}(x)$ is given as $A_{\mu}(x)=\int d^4 x^{\prime}D(x-x^{\prime})J_{\mu}(x^{\prime})$ where $D(x-x^{\prime})$ is the massless retarded propagator. Now $D(x)=\delta(t-r)/4\pi r$ itself is written entirely in terms of the spacetime coordinates and contains no fundamental scales. Rather, the scales reside in the localized $J_{\mu}(x)$, and if it for instance oscillates with a specific frequency, then the resulting $A_{\mu}(x)$ will oscillate with that same frequency too. However, for $J_{\mu}(x)$ to be localized and possess such an oscillation frequency, it would not be scale invariant. Hence the classical Maxwell field only possesses frequency scales because the sources to which it couples are taken to possess them, and the sources themselves can only possess such scales if they are not scale invariant. However, if all the particles contained in electromagnetic sources are to acquire length scales quantum-mechanically, then there could be no fundamental classical electromagnetic sources that could possess such length scales in the first place. Classical electromagnetism with localized oscillating sources is thus a macroscopic manifestation of an underlying microscopic quantum Maxwell theory in which scales are generated dynamically. In a truly scale-free classical electrodynamics there could not be any electromagnetic waves. Thus just like gravity, the same conformal invariance will not permit electromagnetic sources to have any nontrivial intrinsically classical component either, with such sources being intrinsically quantum-mechanical.

If momentum modes are not to arise in classical physics, one needs to ask how it is that they then do arise. To this end we note that as well as generating mass scales via dynamics, there is another way in which quantum mechanics produces scales, namely via the quantization procedure itself. Specifically, scales are introduced via canonical commutation relations, with the generic equal-time commutation relation $[\phi(\bar{x},t),\pi(\bar{x}^{\prime}, t)]=i\hbar \delta^3(\bar{x}-\bar{x}^{\prime})$ for instance being a non-linear relation that introduces a scale $\delta^3(\bar{x}-\bar{x}^{\prime})$ everywhere on a spacelike hypersurface. Since we can set $\delta^3(\bar{x}-\bar{x}^{\prime})=(1/8\pi^3)\int d^3k \exp(i\bar{k}\cdot (\bar{x}-\bar{x}^{\prime}))$, this is equivalent to introducing a complete basis of momentum modes, with momentum modes thus being intrinsically quantum mechanical. And indeed it is the very existence of this set of modes that gives rise to the zero-point energy density and pressure of a quantized field that we discuss in the following. Moreover, a quantized field will have a zero-point energy density and pressure even when it is massless, i.e. even in the absence of  mass generation. Then, when there is mass generation, the momentum modes will obey the $k_0^2=\bar{k}^2+m^2/\hbar^2$ mass condition  and cause the massless theory zero-point energy density and pressure to readjust. And as we show in Secs. (\ref{S7}) and (\ref{S8}), this readjustment will cancel the cosmological constant that is induced by the same mass generation mechanism.

Given the above remarks, we see that in conformal gravity we should expand the metric as a power series in Planck's constant rather than as a power series in the gravitational coupling constant, with the zeroth-order term in the expansion being flat. Consequently, in the theory there is no intrinsic classical gravity, with the equations that are to be used  for macroscopic systems being associated with matrix elements of  the quantum fields in states with an indefinite number of gravitational quanta, Since there is to be no intrinsic classical gravity, there could not be any classical black holes. While conformal gravity thus eliminates the classical gravity singularity problem, and thus simultaneously eliminates the need to have to make such classical singularities compatible with quantum mechanics, it remains to be seen whether the theory might still generate geometric singularities through quantum-mechanical effects, though one might anticipate that the uncertainty principle might spread sources out enough to prevent this from happening.

Even with the requirement that the metric be expanded as a power series in Planck's constant, quantization of gravity can still not follow the standard canonical quantization prescription that is used for other fields. Specifically, for a matter field one obtains its equation of motion by varying the matter action with respect to the matter field, but one obtains its energy-momentum tensor $T^{\mu\nu}_{M}$ by instead varying the matter action with respect to the metric. Since $T^{\mu\nu}_{M}$ involves products of matter fields at the same point, a canonical quantization of the matter field then gives the matter energy-momentum tensor  a non-vanishing zero-point contribution. However, in a standard quantization procedure for a given matter field, the non-vanishing of $T^{\mu\nu}_{M}$ violates no constraint since  one does not simultaneously impose the equation of motion of any other field. Thus for a given matter field one does not require stationarity with respect to the metric, with $T^{\mu\nu}_{M}$ thus not being constrained to vanish. 

In contrast however, for gravity the relevant field is the metric itself. If we define the variation of the gravitational action with respect to the metric to be a quantity $T^{\mu\nu}_{GRAV}$, the gravitational equation of motion is then given by $T^{\mu\nu}_{GRAV}=0$.\footnote{In S. Weinberg, Gravitation and Cosmology: Principles  and Applications of the General Theory of Relativity (Wiley, New York, 1972) and also in P.~D.~Mannheim,~ Phys.~Rev.~D {\bf 74}, 024019 (2006) it is suggested that one treat the functional variation of the gravitational action with respect to the metric as the energy-momentum tensor of gravity. Specifically it was noted that if in Einstein gravity one perturbs the Einstein equations around some background $g^{(0)}_{\mu\nu}$ according to $g_{\mu\nu}=g^{(0)}_{\mu\nu}+h_{\mu\nu}$, the first-order term in $h_{\mu\nu}$  gives the wave equation obeyed by $h_{\mu\nu}$, while the term that is quadratic in $h_{\mu\nu}$ is both covariantly conserved with respect to the background metric and gives the energy density carried by a gravity wave. In Sec. (6) below we provide an equivalent analysis in the conformal case.} Then, with $T^{\mu\nu}_{GRAV}$ containing products of fields at the same point, a canonical quantization of the gravitational field would give a zero-point contribution to $T^{\mu\nu}_{GRAV}$, and thus violate the stationarity condition $T^{\mu\nu}_{GRAV}=0$ that $T^{\mu\nu}_{GRAV}$ has to obey. Hence, unlike the matter fields for which there is no constraint on $T^{\mu\nu}_{M}$ in the absence of any coupling of matter to gravity, gravity itself is always coupled to gravity, with its stationarity condition not permitting it to consistently be quantized on its own. 

Despite this, we note that if we impose a stationarity condition with respect to the metric not on the gravity piece or the matter piece alone, but on their sum as given by the total $I_{\rm UNIV}$ of the universe introduced in (\ref{M9}), we then obtain
\begin{equation}
T^{\mu\nu}_{\rm UNIV}=T^{\mu\nu}_{\rm GRAV}+T^{\mu\nu}_{\rm M}=0.
\label{M20}
\end{equation}
In this case it now is possible to quantize gravity consistently, with $T^{\mu\nu}_{\rm GRAV}$ now being able to be non-zero provided gravity is coupled to some quantized matter field source for which $T^{\mu\nu}_{\rm M}$ is non-zero. Thus gravity can only be quantized consistently if it is coupled to a quantized matter field. However, in order for the cancellation required of the total $T^{\mu\nu}_{\rm UNIV}$ to actually take place, the quantization condition imposed on the gravitational sector commutation relations would have to be fixed by the quantization condition in the matter sector in order to enforce $T^{\mu\nu}_{\rm GRAV}=-T^{\mu\nu}_{\rm M}$, with each term being intrinsically quantum-mechanical. Consequently, gravity is not only quantized though its coupling to quantized matter, its commutation relations are explicitly determined by that coupling, with gravity needing no independent quantization of its own. Finally, we note that not only do the matter fields quantize gravity, the vanishing of $T^{\mu\nu}_{\rm UNIV}$ entails that the gravity field and the matter field zero-point fluctuations must cancel each other identically. In Secs. (\ref{S6}), (\ref{S7}) and (\ref{S8}) we explore this point in detail.

\section{Unitarity via $PT$ symmetry}
\label{S5}

When the $W^{\mu\nu}$ tensor given in (\ref{M12}) is linearized around a flat spacetime background with metric $\eta_{\mu\nu}$ according to $g_{\mu\nu}=\eta_{\mu\nu}+h_{\mu\nu}$, it is found \cite{Mannheim2009a} to be a function of the traceless quantity $K^{\mu\nu}=h^{\mu\nu}-(1/4)\eta^{\mu\nu}\eta_{\alpha\beta}h^{\alpha\beta}$. In the convenient transverse gauge $\partial_{\mu}K^{\mu\nu}=0$, the first order term in $W^{\mu\nu}$ is found to take the simple form
\begin{equation}
W^{\mu\nu}(1)=\frac{1}{2}(\partial_{\alpha}\partial^{\alpha})^2 K^{\mu\nu},
\label{M21}
\end{equation}
while the second order term in the conformal action $I_{\rm W}$ given in (\ref{M9}) takes the form
\begin{equation}
I_{\rm W}(2)=-\frac{\alpha_g}{2}\int d^4x \partial_{\alpha}\partial^{\alpha} K_{\mu\nu}\partial_{\beta}\partial^{\beta} K^{\mu\nu}.
\label{M22}
\end{equation}
Since there is no mixing of components of $K^{\mu\nu}$ in either (\ref{M21}) or (\ref{M22}), one can explore the unitarity structure of the theory by working with an analog one-component scalar field theory. As such, the condition $W^{\mu\nu}(1)=0$ is one of a broad class of fourth-order equations of motion that have been encountered in the literature, and all of them can be associated with the generic scalar action
\begin{equation}
I_{\rm S}=-\frac{1}{2}\int d^4x\left[\partial_{\mu}\partial_{\nu}\phi\partial^{\mu}
\partial^{\nu}\phi+(M_1^2+M_2^2)\partial_{\mu}\phi\partial^{\mu}\phi+M_1^2M^2_2\phi^2\right].
\label{M23}
\end{equation}
Given this action one obtains an equation of motion
\begin{equation}
(-\partial_t^2+\bar{\nabla}^2-M_1^2)(-\partial_t^2+\bar{\nabla}^2-M_2^2)\phi(x)=0,
\label{M24}
\end{equation}
a propagator
\begin{equation}
D(k,M_1,M_2)=\frac{1}{(M_2^2-M_1^2)}\left(\frac{1}{k^2+M_1^2}-\frac{1}{k^2+M_2^2}\right),
\label{M25}
\end{equation}
and an energy-momentum tensor with $(0,0)$ component
\begin{eqnarray}
&&T_{00}(M_1,M_2)
\nonumber\\
&&=\pi_{0}\dot{\phi}+\frac{1}{2}\left[\pi_{00}^2+(M_1^2
+M_2^2)(\dot{\phi}^2-\partial_{i}\phi\partial^{i}\phi)-M_1^2M_2^2\phi^2
-\pi_{ij}\pi^{ij}\right],
\label{M26}
\end{eqnarray}
where 
\begin{eqnarray}
\pi^{\mu}&=&\frac{\partial{\cal L}}{\partial \phi_{,\mu}}-\partial_{\lambda
}\left(\frac{\partial {\cal L}}{\partial\phi_{,\mu,\lambda}}\right)=-(M_1^2+M_2^2)\partial^{\mu}\phi+\partial_{\lambda}\partial^{\mu}\partial^{\lambda}\phi,
\nonumber\\
\pi^{\mu\lambda}&=&\frac{\partial {\cal L}}{\partial \phi_{,\mu,\lambda}}=-\partial^{\mu}\partial^{\lambda}\phi.
\label{M27}
\end{eqnarray}

These equations immediately possess two well-known realizations that exhibit the problems that higher-derivative theories are thought to possess. If one takes the contour for the $k^0$ integration in the propagator to be the standard Feynman one in which all positive energy modes propagate forward in time and all negative energy modes propagate backwards in time, because of the relative minus sign in (\ref{M25}), one finds that some of the poles have the negative residues that occur in an indefinite metric Hilbert space. To avoid such negative residues, one can find \cite{Bender2008b} an alternate contour in which the residues of all poles are positive but in which some of the negative energy modes propagate forward in time. While one can quantize this realization with a standard Dirac norm, as the presence of the $-M_1^2M_2^2\phi^2$ term in $T_{00}(M_1,M_2)$ indicates, in this case the energy eigenvalue spectrum is  unbounded from below.

With neither of these two possibilities being palatable, higher-derivative theories have long been regarded as being unphysical. However, recently Bender and Mannheim revisited the issue \cite{Bender2008a,Bender2008b} and found a third realization of the theory in which the energy spectrum is bounded from below and there are no negative Hilbert space norms at all. With the appropriate scalar product for a Hilbert space being determined by boundary conditions, to determine the relevant scalar product one needs some asymptotic information. To this end Bender and Mannheim studied the eigenvalue problem for the Hamiltonian $H=\int d^3x T_{00}(M_1,M_2)$ in the sector of the theory where the energy eigenvalue spectrum is bounded from below, and found that the associated wave functions were not normalizable on the real axis. In consequence of this, the Hamiltonian of the system could not be Hermitian. However, the wave functions were found to be normalizable on the imaginary axis, and thus the field $\phi$ would have to be anti-Hermitian rather than Hermitian, with the $-M_1^2M_2^2\phi^2$ term in $T_{00}(M_1,M_2)$ then being bounded from below. In addition they noted that if they constructed a path integral for the system, it would not exist with real $\phi$ but would be well-defined if $\phi$ were pure imaginary. So again, one needs to take $\phi$ to be an anti-Hermitian operator. 

Now if a Hamiltonian is not Hermitian, one is immediately concerned that its eigenvalues might not all be real. However, while Hermiticity implies reality of eigenvalues, there is no converse theorem that says that a non-Hermitian Hamiltonian must have complex eigenvalues. Consequently, Hermiticity is only sufficient for reality but not necessary.  Recently, as part of the general $PT$ symmetry program that has been developed by Bender and collaborators \cite{Bender2007} a necessary condition for reality has been found, namely that a Hamiltonian have a $PT$ symmetry where $P$ is a linear operator and $T$ is an antilinear one. Specifically, it was shown in \cite{Bender2002} that if a Hamiltonian is $PT$ invariant the secular equation $|H-\lambda I|=0$ that determines the eigenvalues is real. Then in \cite{Bender2010} the converse was shown, namely that if the secular equation is real, the Hamiltonian must have a $PT$ symmetry. Consequently, the energy eigenspectrum of a Hamiltonian that is not $PT$ symmetric must contain some complex eigenvalues. 

Noting now that all the poles in the propagator given in (\ref{M25}) lie on the real axis, we see that the Hamiltonian for the fourth-order theory while not Hermitian must instead be $PT$ symmetric.  For such Hamiltonians one can construct a norm, the so-called $CPT$ norm of $PT$ theories \cite{Bender2007}, that obeys unitary time evolution. For our purposes here we note that for a non-Hermitian Hamiltonian $H$ that has a completely real energy eigenspectrum, $H$ and $H^{\dagger}$ must be related by a similarity transform of the form 
\begin{equation}
VHV^{-1}=H^{\dagger}.
\label{M28}
\end{equation}
Thus if $H$ has a right-eigenvector according to $H|R\rangle=E|R\rangle$ with real $E$, its conjugate will obey $\langle R|H^{\dagger}=\langle R|E$ and will not be a left-eigenvector of $H$. Rather, the state $\langle L|$ defined as $\langle L|=\langle R|V$ will be a left-eigenvector of $H$ since it obeys
\begin{equation}
\langle L|H=\langle L|E.
\label{M29}
\end{equation}
In this case it will be the norm $\langle L|R\rangle=\langle R|V|R\rangle$ that will obey unitary time evolution since it evolves as 
\begin{equation}
\langle L(t)|R(t)\rangle=\langle L(t=0)|e^{iHt}e^{-iHt}|R(t=0)\rangle=\langle L(t=0)|R(t=0)\rangle. 
\label{M30}
\end{equation}
Thus we see that  the needed scalar product is the overlap of a left-eigenvector with a right-eigenvector and not the overlap of a right-eigenvector with its own conjugate. Moreover, in \cite{Mannheim2009b} it was shown that the existence of a $V$ that can connect $H$ and $H^{\dagger}$ according to $VHV^{-1}=H^{\dagger}$ is a necessary and sufficient condition for both the existence of a $PT$ symmetry and for the existence of a unitary scalar product, with the scalar product then being of the form $\langle L|R\rangle=\langle R|V|R\rangle$. The existence of a $PT$ invariance for a Hamiltonian is thus a necessary and sufficient condition for unitary time evolution.

When these $PT$ ideas are applied to the fourth-order propagator, it is found \cite{Bender2008b,Mannheim2009a} that the relative minus in it is no longer associated with an indefinite metric at all. Rather, it is associated with $V$ operator, with the completeness relation being given by $\Sigma |n\rangle\langle n|V=I$ and not by the negative norm $\Sigma |n\rangle\langle n|-\Sigma |m\rangle\langle m|=I$. Finally, the propagator itself is found to be given by the Green's function $\langle \Omega_L|T(\phi\phi)|\Omega_R\rangle=\langle \Omega_R|VT(\phi\phi)|\Omega_R\rangle$ rather than by the familiar $\langle \Omega_R|T(\phi\phi)|\Omega_R\rangle$. As we see, the unitarity problem for fourth-order propagators only arose because one wanted to represent them as $\langle \Omega_R|T(\phi\phi)|\Omega_R\rangle$. Once one recognizes that the left vacuum need not be the conjugate of the right vacuum unitarity can then readily be achieved. 

Now while the above discussion was developed for the general second- plus fourth-order action given in (\ref{M23}), for the conformal case given in (\ref{M21}) we are interested in a pure fourth-order theory alone where the propagator is given by $D(k)=1/k^4$ (a propagator whose poles again are all on the real axis). Since the reduction to a pure fourth-order theory would require setting both $M_1^2$ and $M_2^2$ equal to zero in (\ref{M23}), we see that because of the $1/(M_2^2-M_1^2)$ prefactor in (\ref{M25}), the limit is singular. In consequence, the limit is a quite unusual one in which the Hamiltonian becomes a non-diagonalizable, Jordan-block Hamiltonian \cite{Bender2008b,Mannheim2009a}, with some of the states that had been eigenstates being replaced by non-stationary ones. The very fact that the Hamiltonian is not diagonalizable immediately confirms that it could not be Hermitian, just as we had noted above. In this case even though the set of energy eigenstates is not complete, the set of stationary plus non-stationary states combined is complete \cite{Bender2008b}, with time evolution of packets built out of the two classes of states combined being unitary \cite{Bender2008b}. The unitarity of the pure fourth-order conformal gravity theory is thus established.

As we thus see, in order to establish unitarity for fourth-order theories we need the field $\phi(x)$, and thus $g_{\mu\nu}(x)$ itself, to be anti-Hermitian rather than Hermitian.  Now this is not how one ordinarily thinks about the gravitational field, since one would presuppose that it, above all fields, should have a real classical limit. Nonetheless, having an anti-Hermitian gravitational field is is not in conflict with anything that is actually known about gravity. Specifically, in \cite{Bender2008b} it was  noted that if one replaces $g_{\mu\nu}$ by $ig_{\mu\nu}$ (and thus $g^{\mu\nu}$ by $-ig^{\mu\nu}$) neither the Christoffel symbols that appear in geodesics nor $R^{\lambda}_{\phantom{\lambda} \mu\nu\sigma}$ would be affected at all. In four space-time dimensions ${\rm det}(g_{\mu\nu})$ would not be affected either. Even though Riemann tensor contractions could generate factors of i, all such factors could be absorbed by redefining the overall multiplicative coefficients in the action (and likewise for the $ds^2=g_{\mu\nu}dx^{\mu}dx^{\nu}$ line element). Hence, current gravitational measurements cannot distinguish between a purely real or a purely imaginary gravitational field. And as we have seen, once one takes the gravitational field to be anti-Hermitian, one can construct  a consistent, renormalizable and unitary theory of quantum gravity. And perhaps the problems that beset quantum gravity have arisen because one wanted the gravitational field to be Hermitian.

\section{The zero-point problem}
\label{S6}

In current applications of standard gravity to macroscopic astrophysical and cosmological systems, one treats gravity itself as being purely classical.  However, one cannot treat its matter source that way too since there are some intrinsically quantum-mechanical sources that are significant macroscopically. Thus, white dwarf stars are stabilized by the Pauli degeneracy pressure of the electrons in the star, and black-body radiation contributes to cosmic expansion.\footnote{The Chandrasekhar mass limit $M_{\rm CH}\sim (\hbar c/G)^{3/2}/m_p^2$ for white dwarfs and the Stefan-Boltzmann constant  $\sigma=2\pi^5k_{\rm B}^4/15c^2\hbar^3$ for black-body radiation both intrinsically depend on $\hbar$. Both of these parameters would expressly have to appear in the macroscopic gravitational equations of motion, and for neither of them could one set $\hbar$ to zero.} To couple these particular effects to classical gravity, both of them are taken to be described as ensemble averages over an appropriate set of positive-energy Fock space states. However, while the Fock space states would be eigenstates of a Hamiltonian of the generic form $\Sigma \hbar \omega (a^{\dagger}a +1/2)$, the infinite $\Sigma \hbar \omega/2$ zero-point contribution is ignored, i.e. one takes the Hamiltonian to be of the truncated form $\Sigma \hbar \omega a^{\dagger}a$ instead. Now in flat space one is free to discard the zero-point term (say by a normal ordering prescription) since in flat space one can only measure energy differences. However the hallmark of gravity is that it couple to energy density itself and not to energy density difference, and so discarding anything to which gravity couples would require justification. 

Since one would have to cancel infinities in the matter field energy-momentum tensor $T^{\mu\nu}_{\rm M}$ if the gravitational effects that  occur in standard gravity are to be finite, some mechanism needs to be identified that would effect the cancellation. An immediate mechanism that might achieve this would be a cancellation between appropriately chosen matter fields, since bosons and fermions contribute to $T^{\mu\nu}_{\rm M}$ with opposite signs. In fact  such a cancellation will occur if there is an exact supersymmetry between fermions and bosons. However, once the fermion-boson mass degeneracy is broken, the cancellation is lost. With the non-observation to date of any of the requisite superpartners, we know that the supersymmetry breaking scale must be at least in the TeV region, with the uncanceled zero-point energy then being huge.

Nonetheless, the generic idea of a boson-fermion cancellation as enforced by some symmetry still makes sense since fermions and bosons generate vacuum energies with opposite signs no matter what the theory. To take advantage of this, we note that the treatment of standard gravity described above is deficient in one rather serious regard, namely it ignores the effect of quantum mechanics on gravity itself. And as soon as one quantizes gravity, gravity itself will acquire a zero-point contribution. Since gravitational quanta are bosonic, under certain circumstances they may then be able to provide the needed cancellation. Specifically, for such a cancellation to occur one needs three things: a quantum gravity theory that makes sense, a symmetry, and the presence of  fermions in the matter field sector. With conformal gravity meeting all of these criteria (as noted above fermions are its building blocks), in the following we will explore its implications for the vacuum energy problem. We will see that when the conformal symmetry is unbroken the needed cancellation does in fact occur. And then, unlike the supersymmetry situation, the cancellation will be maintained even after the conformal symmetry is spontaneously broken and a cosmological constant term is induced. 

One of the challenges that gravity theory faces is that zero-point and cosmological constant  contributions already occur for matter fields in flat spacetime, i.e. they occur in the absence of gravity. Since gravity is not involved in flat space physics, it is very difficult for gravity to then resolve any problem that it is not responsible for. To enable gravity to resolve such problems we need to put the gravitational field an equal footing with the matter fields. This we can do if there are no intrinsic classical contributions in either the gravitational or the matter sectors and all physics is quantum mechanical, i.e. precisely as conformal symmetry requires. Since the lowest order quantum-mechanical contribution to $T^{\mu\nu}_{\rm M}$ is a zero-point contribution of order $\hbar$, to cancel it through the vanishing of the total $T^{\mu\nu}_{\rm UNIV}$ given in (\ref{M20}), we will need the lowest non-trivial gravitational term to be of order $\hbar$ too. Since the zero-point contribution is due to products of fields at the same point,  the order $\hbar$ gravitational zero-point must involve a product of two gravitational fields and thus be given by the second-order $-4\alpha_gW^{\mu\nu}(2)$ ($\equiv T^{\mu\nu}_{\rm GRAV}$) tensor that is obtained by varying the  $I_{\rm W}(2)$ term in (\ref{M22}). To this order in $\hbar$ we only need to evaluate $T^{\mu\nu}_{\rm M}$ in a flat background. It will then generate an order $\hbar$ curvature, with the order $\hbar^2$ term in $T^{\mu\nu}_{\rm M}$ then being curvature dependent. Moreover, with $T^{\mu\nu}_{\rm UNIV}$ vanishing not only in lowest order but in all orders if both the gravitational and matter field sectors are renormalizable (i.e. in a renormalizable theory  (\ref{M20}) is an all-order identity), the zero-point cancellation will occur to all orders. Thus if we decompose $T^{\mu\nu}_{\rm GRAV}$ and $T^{\mu\nu}_{\rm M}$ into finite and divergent parts according to $T^{\mu\nu}_{\rm GRAV}=(T^{\mu\nu}_{\rm GRAV})_{\rm FIN}+(T^{\mu\nu}_{\rm GRAV})_{\rm DIV}$, $T^{\mu\nu}_{\rm M}=(T^{\mu\nu}_{\rm M})_{\rm FIN}+(T^{\mu\nu}_{\rm M})_{\rm DIV}$, (\ref{M20}) will decompose into 
\begin{equation}
(T^{\mu\nu}_{\rm GRAV})_{\rm DIV}+(T^{\mu\nu}_{\rm M})_{\rm DIV}=0,
\label{M31}
\end{equation}
and 
\begin{equation}
(T^{\mu\nu}_{\rm GRAV})_{\rm FIN}+(T^{\mu\nu}_{\rm M})_{\rm FIN}=0.
\label{M32}
\end{equation}
With (\ref{M31}) we see that all gravitational and matter field infinities cancel each other identically, with $(T^{\mu\nu}_{\rm GRAV})_{\rm DIV}$ and $(T^{\mu\nu}_{\rm M})_{\rm DIV}$ regulating each other. Given this regulation, there is no need to renormalize either of the two terms as their sum is finite, and thus no renormalization anomaly such as the conformal anomaly  is generated.\footnote{If one did first renormalize each term separately, the associated conformal anomalies would then have to cancel each other identically. Specifically, with the vanishing of $T^{\mu\nu}_{\rm UNIV}$ being due to stationarity with respect to the metric, such stationarity equally guarantees the  vanishing of the trace $g_{\mu\nu}T^{\mu\nu}_{\rm UNIV}$ without any need to impose conformal invariance. Thus even though the vanishing of the individual gravity sector and matter sector traces $g_{\mu\nu}T^{\mu\nu}_{\rm GRAV}$ and $g_{\mu\nu}T^{\mu\nu}_{\rm M}$ do require conformal invariance, and even though conformal symmetry Ward identities might be violated by renormalization anomalies, the vanishing of $g_{\mu\nu}T^{\mu\nu}_{\rm UNIV}$ cannot be affected by the lack of scale invariance of regulator masses. Any anomalies in $g_{\mu\nu}T^{\mu\nu}_{\rm GRAV}+g_{\mu\nu}T^{\mu\nu}_{\rm M}$ must thus all mutually cancel each other identically.} And with all infinities having been removed, (\ref{M32}) provides us with a completely finite framework for calculating gravitational effects. Thus in (\ref{M31}) we take care of the $\Sigma \hbar \omega/2$ type terms, and in (\ref{M32}) we are free to use the $\Sigma \hbar \omega a^{\dagger}a$ type terms alone.

With $W^{\mu\nu}(2)$ being of order $\hbar$, the gravitational fluctuation $K^{\mu\nu}$ must itself be of order $\hbar^{1/2}$, and since the lowest non-trivial term in $T^{\mu\nu}_{\rm M}$ is of order $\hbar$, it must be the case that $W^{\mu\nu}(1)$ in (\ref{M21}) vanish identically. While the vanishing of $W^{\mu\nu}(1)$ provides us with a wave equation, the situation is not quite the same as the one that occurs when one expands in a power series in the gravitational coupling constant. Specifically, in that case the first-order fluctuation term on the gravitational side is produced by a first-order fluctuation term on the matter side, so that the gravitational fluctuation would obey an inhomogeneous  wave equation with a source. In contrast, in the conformal case the first-order gravitational wave equation is strictly homogeneous on all scales. Then, since this equation is homogeneous, in and of itself it does not force $K^{\mu\nu}$ to be non-zero. However, since the order $\hbar$ contribution to $T^{\mu\nu}_{\rm M}$ is non-zero, $-4\alpha_gW^{\mu\nu}(2)$ cannot vanish, and thus $K^{\mu\nu}$ cannot vanish either. It is thus quantization of the matter field  that forces the gravitational field to be quantized, with the condition $-4\alpha_gW^{\mu\nu}(2)+T^{\mu\nu}_{\rm M}=0$ fixing the strength of the commutator terms present in the second order $W^{\mu\nu}(2)$. With the matter field fixing the strength of the gravitational sector, the cancelation of both zero-point contributions and conformal anomalies is secured.

To see how things work in detail we consider conformal gravity coupled to a Dirac fermion. To the order $\hbar$ of interest to us we can take the fermion to be a free massless fermion in flat spacetime. In this case the matter field energy-momentum tensor is given by $T^{\mu\nu}_{\rm M}=i\hbar \bar{\psi}\gamma^{\mu}\partial^{\nu}\psi$, with its vacuum expectation value being given by 
\begin{equation}
\langle \Omega |T^{\mu\nu}_{\rm M}|\Omega\rangle= -\frac{2\hbar}{(2\pi)^3}\int_{-\infty}^{\infty}d^3k\frac{k^{\mu}k^{\nu}}{\omega_k},
\label{M33}
\end{equation}
where $k^{\mu}$ is a lightlike 4-vector $k^{\mu}=(\omega_k,\bar{k})$ with $\omega_k=|\bar{k}|$.
In (\ref{M33}) we recognize two separate infinite terms, one associated with $\rho_{\rm M}=\langle \Omega |T^{00}_{\rm M}|\Omega\rangle$ and the other with $p_{\rm M}=\langle \Omega |T^{11}_{\rm M}|\Omega\rangle=\langle \Omega |T^{22}_{\rm M}|\Omega\rangle=\langle \Omega |T^{33}_{\rm M}|\Omega\rangle$. Since the fermion is massless, $T^{\mu\nu}_{\rm M}$ is traceless and thus these two infinities obey $\rho_{\rm M}=3p_{\rm M}$. Such an energy-momentum tensor  could not be associated with a cosmological constant term of the form $-\Lambda \eta_{\mu\nu}$ since its trace is given by the non-zero $-4\Lambda$, with the cosmological constant  and zero-point fluctuation problems  in principle thus being different. Given its $k^{\mu}k^{\nu}$ structure, the zero-point quantity $\langle \Omega |T^{\mu\nu}_{\rm M}|\Omega\rangle$ can be written in the form of a perfect fluid with a timelike fluid velocity vector $U^{\mu}=(1,0,0,0)$, viz. 
\begin{equation}
\langle \Omega |T^{\mu\nu}_{\rm M}|\Omega\rangle= (\rho_{\rm M}+p_{\rm M})U^{\mu}U^{\nu}+p_{\rm M}\eta^{\mu\nu},
\label{M34}
\end{equation}
with the fluid thus possessing both a zero-point energy density and a zero-point pressure.\footnote{
Even though (\ref{M33})  involves terms that are infinite and thus not well-defined, we note the perfect fluid form given in (\ref{M34})  can be established by integrating over the direction of the 3-momentum vector $\bar{k}$ alone, an integration that is completely finite. A perfect fluid form for (\ref{M33})  can thus be established prior to the subsequent divergent integration over the magnitude of the momentum, with this latter integration not bringing $\langle \Omega |T^{\mu\nu}_{\rm M}|\Omega\rangle$ to the form of a cosmological constant. Even though (\ref{M33})  is not well-defined, for our purposes here the perfect fluid form given in (\ref{M34}) is a very convenient way of summarizing the infinities in (\ref{M33})  that need to be cancelled.} Since gravity couples to the full $T^{\mu\nu}_{\rm M}$ and not just to its $(0,0)$ component, it is not sufficient to only address the vacuum energy density problem, one has to deal with the vacuum pressure as well. There are thus two vacuum problems that need to be addressed, and not just one. The gravitational sector will thus need to cancel both the vacuum energy density and the vacuum pressure of the matter field, and in a conformal theory will readily be able to do so since $-4\alpha_gW^{\mu\nu}(2)$ has an identical traceless vacuum perfect fluid structure. (In a conformal invariant theory the variation with respect to the metric of the pure gravitational sector of the action is automatically traceless.)

For the explicit structure of the gravity sector we follow the discussion given in \cite{Mannheim2009a}. On using some residual gauge symmetry the general solution to $W^{\mu\nu}(1)=0$ is  given as 
\begin{eqnarray}
K_{\mu\nu}(x)&=&\frac{\hbar^{1/2}}{2(-\alpha_g)^{1/2}}\sum_{i=1}^2\int \frac{d^3k}{(2\pi)^{3/2}(\omega_k)^{3/2}}\bigg{[}
A^{(i)}(\bar{k})\epsilon^{(i)}_{\mu\nu}(\bar{k})e^{ik\cdot x}
\nonumber\\
&+&i\omega_kB^{(i)}(\bar{k})\epsilon^{(i)}_{\mu\nu}(\bar{k})(n\cdot x)e^{ik\cdot x}\
\nonumber\\
&+&\hat{A}^{(i)}(\bar{k})\epsilon^{(i)}_{\mu\nu}(\bar{k})e^{-ik\cdot x}-
i\omega_k\hat{B}^{(i)}(\bar{k})\epsilon^{(i)}_{\mu\nu}(\bar{k})(n\cdot x)e^{-ik\cdot x}\bigg{]},
\label{M35}
\end{eqnarray}
as expressed in terms of quantum operators $A^{(i)}(\bar{k})$, $\hat{A}^{(i)}(\bar{k})$, $B^{(i)}(\bar{k})$ and  $\hat{B}^{(i)}(\bar{k})$ and two transverse traceless polarization tensors $\epsilon^{(i)}_{\mu\nu}(\bar{k})$ ($i=1,2)$, both of which are normalized to $\epsilon_{\alpha\beta}\epsilon^{\alpha\beta}=1$. Since $K^{\mu\nu}$ is to not be Hermitian, the creation operators are not the Hermitian conjugates of the annihilation operators. However, in the following all that will matter is that $A^{(i)}(\bar{k})$ and $B^{(i)}(\bar{k})$ annihilate the right vacuum while $\hat{A}^{(i)}(\bar{k})$ and $\hat{B}^{(i)}(\bar{k})$ annihilate the left vacuum. With $n^{\mu}x_{\mu}$ being equal to $-t$, in (\ref{M35}) we recognize the presence of non-stationary modes, which, as noted above, is characteristic of theories with non-diagonalizable Hamiltonians.

As discussed in detail in \cite{Mannheim2009a}, the quantization procedure is also characteristic of non-diagonalizable Hamiltonians, with the commutators taking the form
\begin{eqnarray}
&&[A^{(i)}(\bar{k}),\hat{B}^{(j)}(\bar{k}^{\prime})]=[B^{(i)}(\bar{k}),\hat{A}^{(j)}(\bar{k}^{\prime})]=Z(k)\delta_{i,j}\delta^3(\bar{k}-\bar{k}^{\prime}),
\nonumber\\
&&[A^{(i)}(\bar{k}),\hat{A}^{(j)}(\bar{k}^{\prime})]=0,\qquad [B^{(i)}(\bar{k}),\hat{B}^{(j)}(\bar{k}^{\prime})]=0,
\nonumber\\
&&[A^{(i)}(\bar{k}),B^{(j)}(\bar{k}^{\prime})]=0,\qquad [\hat{A}^{(i)}(\bar{k}),\hat{B}^{(j)}(\bar{k}^{\prime})]=0.
\label{M36}
\end{eqnarray}
In (\ref{M36}) everything except the possibly $k=|\bar{k}|$ dependent renormalization constant $Z(k)$ is fixed by kinematics (the vanishing of the $[B^{(i)}(\bar{k}),\hat{B}^{(j)}(\bar{k}^{\prime})]$ commutator for instance is needed to cancel all $n^{\mu}x_{\mu}$ terms contained in $-4\alpha_g\langle \Omega|W_{\mu\nu}(2)|\Omega \rangle$). The constant $Z(k)$, however, will be fixed by the dynamics, with the dynamics preventing $Z(k)$ from being zero and the above commutator algebra from being trivial. Specifically, given (\ref{M36}) we obtain
\begin{equation}
-4\alpha_g\langle \Omega|W^{\mu\nu}(2)|\Omega \rangle=\frac{2\hbar}{(2\pi)^3} \int_{-\infty}^{\infty}d^3k\frac{Z(k)k^{\mu}k^{\nu}}{\omega_k},
\label{M37}
\end{equation}
with the factor $2$ appearing in (\ref{M37}) since we have to sum the standard bosonic $\hbar\omega/2$ zero-point energy density per mode over two polarization states of two separate families of massless spin 2 modes (the $A^{(i)}(\bar{k})$ and $B^{(i)}(\bar{k})$ sectors). Then with the fermion sector generating a factor of $-2$ in (\ref{M33}) (the standard fermionic $-\hbar \omega$
zero-point energy density per mode as summed over negative energy states with spin up and spin down) the cancellation of the fermionic and gravitational contributions in $-4\alpha_g\langle \Omega|W^{\mu\nu}(2)|\Omega \rangle+\langle \Omega|T^{\mu\nu}_{\rm M}|\Omega \rangle=0$ will enforce $Z(k)=1$. 

Establishing that $Z$ is fixed by the coupling of gravity to the matter sector is our key result as it shows that gravity requires no independent quantization of its own, with its quantization strength being fixed by the consistency condition that all zero-point infinities cancel identically. To appreciate the point, it is of interest to take a more general matter source. Thus if we take the source to contain $M$ massless gauge bosons and $N$ massless two-component fermions (viz. $N/2$ four-component fermion modes), together they will generate $M-N$ units of $\hbar\omega_k$ for each $\bar{k}$. (For gauge bosons one gets $+\hbar\omega_k/2$ for each of two helicity states.) In this case consistency requires that $Z$ be given by $Z=(N-M)/2$. This condition shows that $Z$ cannot be assigned in isolation. Rather it is determined by the dynamics each time. Moreover since $Z$ must be positive (c.f. no negative norm states) it also provides an interesting constraint on model building, namely that $N$ must be greater than $M$.  For the standard $SU(3)\times SU(2)\times U(1)$ model for instance, we have $M=12$ gauge bosons and $N=16$ two-component spinors per generation, with $Z$ then being positive. Intriguingly, for the grand-unified gauge group $SO(10)$ one has $M=45$ and again $N=16$ per generation, with three generations of fermions thus being the minimum number that would make $Z$ be positive in this case.

A second example of a dynamically determined renormalization constant may be found in two spacetime dimensions ($D=2$). With it being  the Einstein-Hilbert action that is conformally invariant in $D=2$, to the order $\hbar$ of interest to us we thus couple $D=2$ Einstein-Hilbert gravity (with  $1/2\kappa_2^2$ in place of $1/16 \pi G$) to a free $D=2$ massless flat spacetime fermion. Now in $D=2$ the classical Einstein-Hilbert action is a total divergence (the $D=2$ Gauss-Bonnet theorem). Consequently, the associated Feynman path integral is trivial and there is no quantum scattering. However, as noted in \cite{Mannheim2009a,Mannheim2010a}, the theory is not completely empty. Specifically, due to quantum ordering the quantum-mechanical Einstein-Hilbert action is a not a total divergence. (Generically, $A\partial_{\mu}B+B\partial_{\mu}A =\partial_{\mu}(AB)+[B,\partial_{\mu}A]$.) In consequence of this there are zero-point fluctuations in the gravity sector, just as needed to cancel those in the fermion sector.

The specific $D=2$ calculation parallels the $D=4$ case with gravitational fluctuations of the form $g_{\mu\nu}=\eta_{\mu\nu}+h_{\mu\nu}$ being found to satisfy a massless wave equation and with the components of $h_{\mu\nu}$ being related by the vanishing of the trace $\eta_{\mu\nu}h^{\mu\nu}$. With the momentum modes being given by $k^{\mu}=(\omega_k,k)$ where $\omega_k=|k|$, the general solution to the wave equation is given by \cite{Mannheim2009a,Mannheim2010a}
\begin{eqnarray}
h_{00}(x,t)&=&\kappa_2\hbar^{1/2}\int \frac{dk}{(2\pi)^{1/2}(2\omega_k)^{1/2}}\left[A(k)e^{i(kx-\omega_kt)}+C(k)e^{-i(kx-\omega_kt)}\right]
\nonumber \\
&=&h_{11}(x,t),
\nonumber\\
h_{01}(x,t)&=&\kappa_2\hbar^{1/2}\int \frac{dk}{(2\pi)^{1/2}(2\omega_k)^{1/2}}\left[B(k)e^{i(kx-\omega_kt)}+D(k)e^{-i(kx-\omega_kt)}\right].
\nonumber\\
\label{M38}
\end{eqnarray}
On defining 
\begin{eqnarray}
&&\langle \Omega |[C(k),B(k^{\prime})]|\Omega \rangle=-\langle \Omega |B(k)C(k)|\Omega \rangle\delta(k-k^{\prime})=-f_{BC}(k)\delta(k-k^{\prime}),
\nonumber\\
&&\langle \Omega |[A(k),D(k^{\prime})]|\Omega \rangle=~~\langle \Omega |A(k)D(k)|\Omega \rangle\delta(k-k^{\prime})=~~f_{AD}(k)\delta(k-k^{\prime}),
\nonumber\\
\label{M39}
\end{eqnarray}
where $k$ is the spatial component of $k^{\mu}$, which can be positive or negative, we find that the order $\hbar$ vanishing of $T^{\mu\nu}_{\rm GRAV}+T^{\mu\nu}_{\rm M}$ then leads to the condition
\begin{equation}
k[f_{BC}(k)-f_{AD}(k)]=4\omega_k=4|k|,
\label{M40}
\end{equation}
with $f_{BC}(k)-f_{AD}(k)$ being an odd function of $k$. As we see, the matter sector has again fixed the commutation relations for the gravitational field.

\section{The cosmological constant problem in $D=2$}
\label{S7}

In dynamical generation of fermion masses one has to change the vacuum from the normal one $|N\rangle$ in which $\langle N|\bar{\psi}\psi|N\rangle$  is zero to a spontaneously broken one $|S\rangle$ in which $\langle S|\bar{\psi}\psi|S\rangle$ is non-zero. Since (\ref{M20}) is an operator identity it will hold in any state, and thus the cancellations required to enforce $\langle S|T^{\mu\nu}_{\rm GRAV}|S \rangle+\langle S|T^{\mu\nu}_{\rm M}|S \rangle=0$ must occur. To see how this explicitly comes about, it is instructive to consider a four-Fermi interaction in two spacetime dimensions, as that is the dimension in which the four-Fermi coupling constant $g$ is dimensionless and the theory is conformal invariant. We introduce a flat spacetime four-Fermi action of the form $I_{\rm M}=-\int d^2x[i\hbar\bar{\psi}\gamma^{\mu}\partial_{\mu}\psi-(g/2)(\bar{\psi}\psi)^2]$, with the energy-momentum tensor $T^{\mu\nu}_{\rm M}=i\hbar\bar{\psi}\gamma^{\mu}\partial^{\nu}\psi -\eta^{\mu\nu}(g/2)[\bar{\psi}\psi]^2$ being traceless in solutions to the equation of motion, just as must be the case for conformal matter.

In the Nambu-Jona-Lasinio mean-field, Hartree-Fock approximation one looks for self-consistent states $|S\rangle$ in which  $\langle S|\bar{\psi}\psi | S\rangle=im/g$ and $\langle S|(\bar{\psi}\psi -im/g)^2| S\rangle=0$. In such states the fermion equation of motion takes the form $i\hbar\gamma^{\mu}\partial_{\mu}\psi -im\psi=0$ and the mean-field energy-momentum tensor $T^{\mu\nu}_{\rm MF}$  takes the form 
\begin{equation}
\langle S|T^{\mu\nu}_{\rm MF}| S\rangle=\langle S|i\hbar\bar{\psi}\gamma^{\mu}\partial^{\nu}\psi| S\rangle +\frac{m^2}{2g}\eta^{\mu\nu}, 
\label{M41}
\end{equation}
with the mean-field approximation preserving tracelessness. With the fermion momentum modes being given by $k^{\mu}=(\omega_k,k)$ where $\omega_k=(k^2+m^2/\hbar^2)^{1/2}$, the quantity $\langle S|i\hbar\bar{\psi}\gamma^{\mu}\partial^{\nu}\psi| S\rangle$ evaluates to 
\begin{equation}
\langle S|i\hbar\bar{\psi}\gamma^{\mu}\partial^{\nu}\psi| S\rangle= -\frac{\hbar}{2\pi}\int_{-\infty}^{\infty}dk\frac{k^{\mu}k^{\nu}}{\omega_k}.
\label{M42}
\end{equation}
In (\ref{M42}) we recognize mean-field energy density and pressure terms of the form 
\begin{eqnarray}
\rho_{\rm MF}&=&-\frac{\hbar}{2\pi}\left[K^2+\frac{m^2}{2\hbar^2}+\frac{m^2}{2\hbar^2}{\rm ln}\left(\frac{4 \hbar^2K^2}{m^2}\right)\right],
\nonumber\\
p_{\rm MF}&=&-\frac{\hbar}{2\pi}\left[K^2+\frac{m^2}{2\hbar^2}-\frac{m^2}{2\hbar^2}{\rm ln}\left(\frac{4 \hbar^2K^2}{m^2}\right)\right],
\label{M43}
\end{eqnarray}
as conveniently cut-off at a momentum $K$ that serves to characterize the infinities involved. In  the $(m^2/2g)\eta^{\mu\nu}$ term in (\ref{M41}) we recognize a mean-field cosmological constant term $\Lambda_{\rm MF} =-m^2/2g$, and with $\Lambda_{\rm MF}$ evaluating to the logarithmically divergent 
\begin{equation}
\Lambda_{\rm MF} =\frac{m^2}{4\pi\hbar}{\rm ln}\left(\frac{4 \hbar^2K^2}{m^2}\right),
\label{M44}
\end{equation}
we obtain the gap equation $m=2 \hbar Ke^{\pi\hbar/g}$. (In dynamical symmetry breaking the induced cosmological constant is infinite rather than finite -- it thus appears in $(T^{\mu\nu}_{\rm M})_{\rm DIV}$ and not in $(T^{\mu\nu}_{\rm M})_{\rm FIN}$.) In terms of  $\rho_{\rm MF}$, $p_{\rm MF}$ and $\Lambda_{\rm MF}$ we can write the complete mean-field $\langle S|T^{\mu\nu}_{\rm MF}| S\rangle$ as
\begin{equation}
\langle S|T^{\mu\nu}_{\rm MF}| S\rangle=(\rho_{\rm MF}+p_{\rm MF})U^{\mu}U^{\nu}+p_{\rm MF}\eta^{\mu\nu} -\Lambda_{\rm MF}\eta^{\mu\nu}.
\label{M45}
\end{equation}

Since $\langle S|T^{\mu\nu}_{\rm MF}| S\rangle$ is traceless, the various terms in (\ref{M45}) must obey $p_{\rm MF}-\rho_{\rm MF} -2\Lambda_{\rm MF}=0$ (in $D=2$), with all the various divergences canceling each other in the trace, just as noted in \cite{Mannheim2008,Mannheim2009a}. Given this cancellation, we can eliminate $\Lambda_{\rm MF}$ and rewrite (\ref{M45}) in the manifestly traceless form 
\begin{eqnarray}
\langle S|T^{\mu\nu}_{\rm MF}| S\rangle&=&\frac{(\rho_{\rm MF}+p_{\rm MF})}{2}\left[2U^{\mu}U^{\nu}+\eta^{\mu\nu}\right],
\nonumber\\
\frac{\rho_{\rm MF}+p_{\rm MF}}{2}&=&\langle S|T^{00}_{\rm MF}| S\rangle
=-\frac{\hbar}{2\pi}\left(K^2+\frac{m^2}{2\hbar^2}\right),
\label{M46}
\end{eqnarray}
with the logarithmic divergences associated with the readjustment of $\rho_{\rm MF}$ and $p_{\rm MF}$ in (\ref{M43}) from the massless to the massive case having completely disappeared.  Finally, in order for gravity to now cancel $\langle S|T^{\mu\nu}_{\rm MF}| S\rangle$, we have to replace (\ref{M40}) by
\begin{equation}
k[f_{BC}(k)-f_{AD}(k)]=4\left[(k^2+m^2/\hbar^2)^{1/2}-\frac{m^2}{2\hbar^2(k^2+m^2/\hbar^2)^{1/2}}\right].
\label{M47}
\end{equation}
Thus in the presence of dynamical symmetry breaking, even though the gravitational sector modes remain massless, the commutator renormalization constants in  (\ref{M39}) readjust and become dependent on the induced fermion mass, with the renormalization constants thus being dependent on the choice of vacuum. The emergence of a behavior such as this is completely foreign to the standard canonical commutation prescription used for the matter fields, and shows how different quantization for gravity has to be. Nonetheless, with this readjustment, the vacuum cosmological constant term is completely cancelled, with conformal gravity thus being able to control the cosmological constant even after the conformal symmetry is broken.

\section{The cosmological constant problem in $D=4$}
\label{S8}

To generalize the $D=2$ results to $D=4$ is not immediate  since in $D=4$ the four-Fermi interaction is not conformal invariant.  Rather one must work in a conformal invariant theory, which in $D=4$ means a gauge theory. Since the renormalization procedure would introduce scaling anomaIies, to restore scale symmetry to the gauge theory one needs to be at a renormalization group fixed point. In such a case scaling would be restored but with anomalous dimensions, something first noted by Johnson, Baker and Wiley \cite{Johnson1964} in a study of (flat spacetime) quantum electrodynamics at a Gell-Mann-Low eigenvalue for the fine structure constant. In a study of dynamical symmetry breaking in this same theory it was found  \cite{Mannheim1974} if the dimension $d_{\theta}=3+\gamma_{\theta}$ of the fermion composite bilinear $\theta=\bar{\psi}\psi$ is reduced by one whole unit from its canonical value of $d_{\theta}=3$ to an anomalous value of $d_{\theta}=2$, the vacuum would then undergo dynamical symmetry breaking and generate a fermion mass. Specifically, with the insertion of $\bar{\psi}\psi$ into the inverse fermion propagator behaving as $\tilde{\Gamma}_{\theta}(p,p,0)=(-p^2/M^2)^{-1/2}$ at $\gamma_{\theta}=-1$, it was found that the four-Fermi value of $\epsilon(m,{\rm 4F})=(i/\hbar)\int d^4p/(2\pi)^4{\rm Tr Ln}(\gamma^{\mu}p_{\mu}-m+i\epsilon)$, viz.
\begin{equation}
\epsilon(m,{\rm 4F})=-\frac{\hbar}{4\pi^2}\left(K^4 +\frac{m^2K^2}{\hbar^2} -{m^4 \over 4\hbar^4}{\rm ln}\left({4\hbar^2K^2 \over m^2}\right)+{m^4\over 8\hbar^4}\right),
\label{M48}
\end{equation}
would change to 
\begin{eqnarray}
\epsilon(m)&=&\frac{i}{\hbar}\int \frac{d^4p}{(2\pi)^4}{\rm Tr Ln}\left[\gamma^{\mu}p_{\mu}-m\left(\frac{-p^2}{M^2}\right)^{-1/2}+i\epsilon\right]
\nonumber\\
&=&-\frac{\hbar K^4}{4\pi^2}+ \frac{m^2M^2}{16\pi^2\hbar^3}\left[{\rm ln}\left(\frac{m^2M^2}{16\hbar^4K^4}\right)-1\right].
\label{M49}
\end{eqnarray}
On setting  $\epsilon^{\prime}(m)=m/g$ and $M^4=16\hbar^4K^4{\rm exp}(8\pi^2\hbar^3/M^2g)$, the mean-field energy density  $\epsilon(m)-m^2/2g$ thus evaluates to 
\begin{equation}
\epsilon(m)-\frac{m^2}{2g}=-\frac{\hbar K^4}{4\pi^2}+ \frac{m^2M^2}{16\pi^2\hbar^3}\left[{\rm ln}\left(\frac{m^2}{M^2}\right)-1\right]. 
\label{M50}
\end{equation}
Other than the $m$-independent quartically divergent term (which also occurs in the gravity sector) the mean-field energy density is completely finite, with a local maximum at $m=0$ and global minima at $m=\pm M$. With the four-Fermi $\epsilon(m,{\rm 4F})$ given in (\ref{M48}) having mass-dependent terms that are quadratically and logarithmically divergent, we see that improving the short distance-behavior of $\bar{\psi}\psi$ by one whole unit, and thus that of $\bar{\psi}\psi \bar{\psi}\psi$ by two whole units, then brings the quadratic divergence down to logarithmic; with the $-m^2/2g$ term then removing the logarithmic divergence, to produce the finite terms in 
(\ref{M50}).\footnote{It was noted in \cite{Mannheim1974} that a reduction in dynamical dimension of $\bar{\psi}\psi \bar{\psi}\psi$ from six to four would render the Nambu-Jona-Lasinio four-Fermi interaction theory non-perturbatively renormalizable. In addition it was suggested that one could use such a $\bar{\psi}\psi \bar{\psi}\psi$ term as vacuum energy  counter-term. Now since the QED study given in \cite{Mannheim1974}  was a purely flat spacetime study, there one could of course, and indeed one ordinarily does,  remove the vacuum energy density by normal ordering. However, once one couples the theory to gravity, one can no longer normal order away any contribution to the energy density, and to obtain a finite vacuum energy density one should instead use a four-Fermi counter-term. The $d_{\theta}=2$ condition is thus seen as not only serving to produce dynamical symmetry breaking in flat spacetime, but also as serving to render zero-point fluctuations finite in curved spacetime.}

At the minimum, (\ref{M50}) takes the form
\begin{equation}
\epsilon(M)-\frac{M^2}{2g}=-\frac{\hbar K^4}{4\pi^2}- \frac{M^4}{16\pi^2\hbar^3},
\label{M51}
\end{equation}
in complete analog to (\ref{M46}). As required by (\ref{M20}), gravity must thus cancel the whole of (\ref{M51}) in $D=4$ just as it cancels the whole of (\ref{M46}) in $D=2$. To this end we note that the propagator $S(p)=[\gamma^{\mu}p_{\mu}-m(-p^2/M^2)^{-1/2}+i\epsilon]^{-1}$ contained in (\ref{M49}) has poles, and they can be taken to be at  $p^4-m^2M^2=0$ if we define the multiple-valued square root singularity appropriately. If we do the $p^0$ contour integration in (\ref{M49}) we will obtain poles at $p^2=mM$ and $p^2=-mM$. Recalling that $\langle S|\bar{\psi}\psi |S\rangle=\epsilon^{\prime}(m)$, in analog to (\ref{M33}) we can set  
\begin{eqnarray}
\epsilon(m)-\frac{m^2}{2g}&=&-\frac{2\hbar}{(2\pi)^3}\int_{-\infty}^{\infty}d^3k
\bigg{[}(k^2+mM/\hbar^2)^{1/2}-\frac{mM}{4\hbar^2(k^2+mM/\hbar^2)^{1/2}} 
\nonumber\\
&+&(k^2-mM/\hbar^2)^{1/2}+\frac{mM}{4\hbar^2(k^2-mM\hbar^2)^{1/2}}\bigg{]}.
\label{M52}
\end{eqnarray}
Thus at the $m=M$ minimum we obtain
\begin{eqnarray}
\epsilon(M)-\frac{M^2}{2g}&=&-\frac{2\hbar}{(2\pi)^3}\int_{-\infty}^{\infty}d^3k
\bigg{[}(k^2+M^2/\hbar^2)^{1/2}-\frac{M^2}{4\hbar^2(k^2+M^2/\hbar^2)^{1/2}} 
\nonumber\\
&+&(k^2-M^2/\hbar^2)^{1/2}+\frac{M^2}{4\hbar^2(k^2-M^2/\hbar^2)^{1/2}}\bigg{]}.
\label{M53}
\end{eqnarray}
Comparing now with (\ref{M37}), we see that the cancellation of the fermionic and gravitational contributions in $-4\alpha_g\langle S|W^{\mu\nu}(2)|S \rangle+\langle S|T^{\mu\nu}_{\rm M}|S \rangle=0$ will force the gravitational sector renormalization constant $Z(k)$ in (\ref{M36}) to obey
\begin{eqnarray}
kZ(k)&=&(k^2+M^2/\hbar^2)^{1/2}-\frac{M^2}{4\hbar^2(k^2+M^2/\hbar^2)^{1/2}} 
\nonumber\\
&+&(k^2-M^2/\hbar^2)^{1/2}+\frac{M^2}{4\hbar^2(k^2-M^2/\hbar^2)^{1/2}},
\label{M54}
\end{eqnarray}
in complete analog to (\ref{M47}), with $Z(k)$ again being determined by the dynamics. (In  (\ref{M47}) and (\ref{M54}) the numerical factor of $2$ or $4$ factor in the denominator is the spacetime dimension.) Additionally we note that if we were to set $M=0$ in (\ref{M54}) we would obtain $Z(k)=2$ rather than $Z(k)=1$, since the pole structure of the propagator $S(p)$ is that of two 4-component fermions rather than just one.

Having now obtained (\ref{M53}) and (\ref{M54}), we note that there is an alternate way to derive the structure given in (\ref{M53}) and (\ref{M54}) that is instructive in its own right. Since we are in a conformal theory we can treat the two sets of poles at $p^2=M^2$ and $p^2=-M^2$ as though they were independent degrees of freedom each with the traceless energy-tensor $T_{\mu\nu}=i\bar{\psi}\gamma_{\mu}\partial_{\nu}\psi -(1/4)\eta_{\mu\nu}\bar{\psi}\psi$ required in the broken symmetry case \cite{Mannheim2006}. Recognizing $T^{00}=i\bar{\psi}\gamma^{0}\partial^{0}\psi -(1/4)\eta^{00}\bar{\psi}\psi$ to be in the generic form $T^{00}=\epsilon(m)-(m/4)d\epsilon(m)/dm$ for a particle of mass $m$, we recognize $\epsilon(M)-(M/4)d\epsilon(M)/d(M)+\epsilon(iM)-(iM/4)d\epsilon(iM)/d(iM)$ as being none other than the right-hand side of (\ref{M53}).

\section{The dark matter problem}
\label{S9}

Since conformal gravity is a well-defined, renormalizable quantum theory, we can take matrix elements of (\ref{M32}) in states with an indefinite number of gravitational quanta and obtain a completely finite macroscopic limit that will be described by a classical version of (\ref{M11}). Classical conformal gravity has been studied by Mannheim and Kazanas \cite{Mannheim1989} who found that because of the underlying conformal symmetry,  the exact metric in a static, spherically symmetry geometry can be brought to the form
\begin{equation}
ds^2=-B(r)dt^2+\frac{dr^2}{B(r)}+r^2d\Omega_2,
\label{M55}
\end{equation}
where the metric coefficient $B(r)$ obeys the the fourth-order equation
\begin{eqnarray}
\frac{3}{B(r)}(W^{0}_{\phantom{0}0}-W^{r}_{\phantom{r}r})&=&\nabla^4B=B^{\prime\prime\prime\prime}+\frac{4B^{\prime\prime\prime}}{r}=
\frac{(rB)^{\prime\prime\prime\prime}}{r} 
\nonumber\\
&=&\frac{3}{4\alpha_gB(r)}(T^{0}_{\phantom{0}0}-T^{r}_{\phantom{r}r})\equiv f(r)
\label{M56}
\end{eqnarray}
without any approximation whatsoever. Exterior to a source of radius $r_0$ the solution to (\ref{M56}) is of the form 
\begin{equation}
B(r>r_0)=1-\frac{2\beta}{r}+\gamma r ,
\label{M57}
\end{equation}
with the matching of the interior and exterior solutions fixing the integration constants in (\ref{M57}) according to
\begin{equation}
2\beta=\frac{1}{6}\int_0^{r_0}dr^{\prime}r^{\prime 4}f(r^{\prime}),\qquad \gamma= -\frac{1}{2}\int_0^{r_0}dr^{\prime}r^{\prime 2}f(r^{\prime}) .
\label{M58}
\end{equation}
Comparing with (\ref{M5}) and (\ref{M6}) we see that, as had been been noted above, we do indeed recover the Schwarzschild solution, but in addition  we see that we obtain a linear potential, to thus give a departure from Newton-Einstein at large distances, i.e. in precisely the region where the dark matter problem is encountered.

Given (\ref{M57}), we see that a star would put out a weak gravity potential 
\begin{equation}
V^{*}(r)=-\frac{\beta^{*}c^2}{r}+\frac{\gamma^{*} c^2 r}{2}
\label{M59}
\end{equation}
per unit solar mass. In spiral galaxies the luminous matter at a radial distance $R$ from the galactic center is typically distributed with a surface brightness $\Sigma (R)=\Sigma_0e^{-R/R_0}$, with the total luminosity being given by $L=2\pi \Sigma_0R_0^2$. If we assume that the mass distribution in a spiral galaxy is the same as that of its luminous distribution (i.e. no dark matter), then for a galactic mass to light ratio $M/L$, one can define the total number of solar mass units $N^{*}$ in the galaxy via $(M/L)L=M=N^{*}M_{\odot}$. On integrating $V^{*}(r)$ over this visible matter distribution, one finds that the net centripetal acceleration due to the local luminous matter in the galaxy is given by \cite{Mannheim2006}   
\begin{eqnarray}
\frac{v_{{\rm LOC}}^2}{R}&=&
\frac{N^*\beta^*c^2 R}{2R_0^3}\left[I_0\left(\frac{R}{2R_0}
\right)K_0\left(\frac{R}{2R_0}\right)-
I_1\left(\frac{R}{2R_0}\right)
K_1\left(\frac{R}{2R_0}\right)\right]
\nonumber \\
&&+\frac{N^*\gamma^* c^2R}{2R_0}I_1\left(\frac{R}{2R_0}\right)
K_1\left(\frac{R}{2R_0}\right).
\label{M60}
\end{eqnarray} 

Familiarity with Newtonian gravity would suggest that to fit galactic rotation curve data in conformal gravity one should now apply (\ref{M60}) as is. However there is a crucial difference between the two cases. For Newtonian gravity one  uses the second-order Poisson equation $\nabla^2\phi(r)=g(r)$ and obtains a potential and force of the form 
\begin{equation}
\phi(r)= -\frac{1}{r}\int_0^r
dr^{\prime}r^{\prime 2}g(r^{\prime})-\int_r^{\infty}
dr^{\prime}r^{\prime }g(r^{\prime}),\qquad \frac{d\phi(r)}{dr}= \frac{1}{r^2}\int_0^r
dr^{\prime}r^{\prime 2}g(r^{\prime}).
\label{M61}
\end{equation}                                 
As such, the import of (\ref{M61}) is that even though $g(r)$ could continue globally all the way to infinity, the force at any radial point $r$ is determined solely by the material in the local $0< r^{\prime}< r$ region. In this sense Newtonian gravity is local in character, since to explain a gravitational effect in some local region one only needs to consider the material in that region. Thus in Newtonian gravity, if one wishes to explain the behavior of galactic rotation curves through the use of dark matter, one must locate the dark matter where the problem is and not elsewhere, i.e. within the galaxies themselves.

However, this familiar property of Newtonian gravity is not generic to any theory of gravity. In particular if we define
$h(r)=c^2f(r)/2$, the conformal gravity potential associated with (\ref{M56}) will obey the fourth-order Poisson equation $\nabla^4\phi(r)=h(r)$, with general solution
\begin{eqnarray}
\phi(r)&=&-\frac{r}{2}\int_0^r
dr^{\prime}r^{\prime 2}h(r^{\prime})
-\frac{1}{6r}\int_0^r
dr^{\prime}r^{\prime 4}h(r^{\prime})
-\frac{1}{2}\int_r^{\infty}
dr^{\prime}r^{\prime 3}h(r^{\prime})
\nonumber\\
&&-\frac{r^2}{6}\int_r^{\infty}
dr^{\prime}r^{\prime }h(r^{\prime})
\nonumber \\
\frac{d\phi(r)}{dr}&=&-\frac{1}{2}\int_0^r
dr^{\prime}r^{\prime 2}h(r^{\prime})
+\frac{1}{6r^2}\int_0^r
dr^{\prime}r^{\prime 4}h(r^{\prime})
-\frac{r}{3}\int_r^{\infty}
dr^{\prime}r^{\prime }h(r^{\prime}).
\label{M62}
\end{eqnarray}                                 
As we see, this time we do find a global contribution to the force coming from material in the $r< r^{\prime}< \infty $ region that is beyond the radial point of interest. Hence in conformal gravity one cannot ignore the rest of the universe, with a test particle in orbit in a galaxy being able to sample both the local field due to the matter in the galaxy and the global field due to the rest of the matter in the Universe. Unlike Newtonian gravity then, conformal gravity is an intrinsically global theory.

The contribution that the rest of the Universe provides consists of two components, the homogeneous cosmological background and the inhomogeneities within it. The homogeneous background can be described by a Roberston-Walker (RW) geometry, while large scale inhomogeneities are typically in the form of large gravitationally bound systems such as clusters and superclusters. Since the RW metric is conformal to flat, and since the Weyl tensor vanishes identically in a such a geometry, the cosmological background is characterized by a geometry in which $W^{\mu\nu}$ of (\ref{M11}) (and thus the cosmological $T^{\mu\nu}_{\rm M}$) vanish identically. However, since localized inhomogeneities have a non-vanishing Weyl tensor, the inhomogeneities contribute to the integrals in (\ref{M62}) that extend out to infinity beyond the galaxy of interest. The inhomogeneities contribute to the particular integral solution to (\ref{M11})  given in (\ref{M62}) (both $\nabla^4B(r)$ and $f(r)$ non-zero), while the homogeneous background contributes to the complementary function (both $\nabla^4B(r)$ and $f(r)$ zero).

In order for the background cosmology to contribute non-trivially, we note that even though we need the background $T^{\mu\nu}_{\rm M}$ to vanish (since the RW $W^{\mu\nu}$ vanishes), we would need $T^{\mu\nu}_{\rm M}$ to vanish non-trivially if it is to have any content. As shown in \cite{Mannheim2006}, such a non-trivial vanishing can be achieved by an interplay between the positive contribution of the matter sources (c.f. $\Sigma \hbar \omega a^{\dagger}a$) and the negative contribution of the gravitational field that occurs if the 3-curvature $K$ of the Universe is negative, with gravity providing negative energy density. In \cite{Mannheim2006} it was shown that with such a cosmology one could then fit the accelerating universe Hubble plot data without the need for any fine-tuning of parameters or for any of the cosmological dark matter required in the standard theory. (Unlike the standard $\Omega_{\rm K}=0$ cosmology, which is fine-tuned to only accelerate at late redshift, with its negative $K$ the conformal cosmology naturally accelerates at all redshifts, to thereby lead \cite{Mannheim2006} to a non fine-tuned fit to the acccelerating Universe supernovae Hubble plot data.)

Since cosmology is written in comoving Hubble flow coordinates while rotation curves are measured in galactic rest frames, to ascertain the impact of cosmology on rotation curves one needs to transform the RW metric to static coordinates. As noted in \cite{Mannheim1989}, the transformation
\begin{equation}
\rho=\frac{4r}{2(1+\gamma_0r)^{1/2}+2 +\gamma_0 r},\qquad \tau=\int dt R(t)
\label{M63}
\end{equation}                                 
effects the metric transformation
\begin{eqnarray}
&&-(1+\gamma_0r)c^2dt^2+\frac{dr^2}{(1+\gamma_0r)}+r^2d\Omega_2=
\nonumber \\
&&\frac{1}{R^2(\tau)}\left(\frac{1+\gamma_0\rho/4}
{1-\gamma_0\rho/4}\right)^2
\left[-c^2d\tau^2+\frac{R^2(\tau)}{[1-\gamma_0^2\rho^2/16]^2}
\left(d\rho^2+\rho^2d\Omega_2\right)\right].
\label{M64}
\end{eqnarray} 
Recognizing (\ref{M64}) to be conformal to a topologically open RW metric with 3-curvature $K=-\gamma_0^2/4$, and recalling that in metrics conformal to RW the tensor $W^{\mu\nu}$ still vanishes, we see that in the rest frame of a galaxy the negative $K$ global cosmology found in \cite{Mannheim2006} acts like a universal linear potential with cosmological strength $\gamma_0/2=(-K)^{1/2}$. 

In the weak gravity limit one can add this global potential on to (\ref{M60}), with the total centripetal acceleration then being given by \cite{Mannheim1997} 
\begin{equation}
\frac{v^2_{\rm TOT}}{R}=\frac{v^2_{\rm LOC}}{R}+\frac{\gamma_0 c^2}{2}.
\label{M65}
\end{equation}                                 
In \cite{Mannheim1997}  (\ref{M65}) was used to fit the galactic rotation velocities of a sample of 11 spiral galaxies, and good fits were found, with the two universal linear potential parameters being fixed to the values 
\begin{equation}
\gamma^*=5.42\times 10^{-41} {\rm cm}^{-1},\qquad \gamma_0=3.06\times
10^{-30} {\rm cm}^{-1}.
\label{M66}
\end{equation} 
The value obtained for $\gamma^*$ entails that the linear potential of the Sun is so small that there are no modifications to standard solar system phenomenology, with the values obtained for $N^*\gamma^*$ and $\gamma_0$ being such that one has to go to galactic scales before their effects can become as big as the Newtonian contribution. 

However, as we had noted above, there is a contribution due to inhomogeneities in the cosmic background that we need to include too. These inhomogeneities would typically be clusters and superclusters and would be associated with distance scales between 1 Mpc and 100 Mpc or so. Without knowing anything other than that about them, we see from (\ref{M62}) that  for calculating potentials at galactic distance scales (viz. scales much less than cluster scales of order $r_{\rm clus}$) the inhomogeneities would contribute constant and quadratic terms multiplied by integrals that are evaluated between end points such as $r_{\rm clus}$ that do not depend on the galaxy of interest, to thus be constants.  Thus, as noted in \cite{Mannheim2010b,Mannheim2010c,O'Brien2011}, we augment (\ref{M65}) to 
\begin{equation}
\frac{v^2_{\rm TOT}}{R}=\frac{v^2_{\rm LOC}}{R}+\frac{\gamma_0 c^2}{2}-\kappa c^2R,
\label{M67}
\end{equation}                                 
with asymptotic limit 
\begin{equation}
\frac{v_{{\rm TOT}}^2}{R} \rightarrow \frac{N^*\beta^*c^2}{R^2}+
\frac{N^*\gamma^*c^2}{2}+\frac{\gamma_0c^2}{2}-\kappa c^2R,
\label{M68}
\end{equation} 
where $\kappa=(1/3c^2)\int _{r_{\rm clus}}^{\infty} dr^{\prime}r^{\prime}h(r^{\prime})$. Armed with (\ref{M67}) Mannheim and O'Brien \cite{Mannheim2010b,Mannheim2010c,O'Brien2011} set out to extend the earlier 11 galaxy study of \cite{Mannheim1997} to a total sample of 138 galaxies that had become available in the interim (a varied and broad sample that includes both high and low surface brightness galaxies and dwarfs). In making such fits the only parameter that can vary from one galaxy to the next is the galactic disk mass to light ratio as embodied in $N^*$, with the parameters $\gamma^{*}$, $\gamma_0$ and $\kappa $ needing to be universal and not have any dependence on a given galaxy at all. To model the contribution of the luminous matter known photometric surface brightness data parameters and HI gas data parameters were used. The fits are thus highly constrained, one parameter per galaxy fits with all input data being known, with everything else being universal, and with no dark matter being assumed.

Now since the $\kappa$-dependent term had not been used in the fits given in \cite{Mannheim1997}, one would immediately expect that it would be too small to be significant. However, because the 138 galaxy sample is so big, it contains galaxies whose rotation velocity data go out to radial distances much larger than the ones that had previously been considered. These data are thus sensitive to the distance-dependent $-\kappa c^2R$ term present in (\ref{M67}), with the fitting underscoring the value of working with a large data sample. The fitting to the complete 138 galaxy sample is reported in \cite{Mannheim2010c,O'Brien2011}, with the fitting to the 21 largest galaxies (viz. those that are most sensitive to the  $-\kappa c^2R$ term) being reported here and in  \cite{Mannheim2010b}. The fitting shows that without any galactic dark matter (\ref{M67}) captures the essence of the data for the entire 138 galaxy sample, with the parameters $\gamma^*$ and $\gamma_0$ continuing to take the values given in (\ref{M66}), and with $\kappa$ being found to take a typical  cluster-sized value
\begin{equation}
\kappa =9.54\times 10^{-54}~{\rm cm}^{-2}\approx (100~{\rm Mpc})^{-2}.
\label{M69}
\end{equation} 

In the figures we present the actual fitting to the 21 galaxy sample with all details being given in \cite{Mannheim2010b,Mannheim2010c,O'Brien2011}. In the figures the rotational velocities and errors  (in ${\rm km}~{\rm sec}^{-1}$) are plotted as a function of radial distance (in ${\rm kpc}$). For each galaxy we exhibit the contribution due to the luminous Newtonian term alone (dashed curve), the contribution from the two linear terms alone (dot dashed curve), the contribution from the two 
linear terms and the quadratic terms combined (dotted curve), with the full curve showing the total contribution. Because the data go out to such large distances the data are sensitive to the rise in velocity associated with the linear potential terms, and it is here that the quadratic term acts to actually arrest  the rise altogether (dotted curve) and cause all rotation velocities to ultimately fall. Moreover, since $v^2$ cannot be negative, beyond a distance $R$ of order $\gamma_0/\kappa  =3.21\times 10^{23}~{\rm cm}$ or so there could no longer be any bound galactic orbits, with galaxies thus having a natural way of terminating, and with global physics thus imposing a natural limit on the size of galaxies. To illustrate this we plot the rotation velocity curves for the galaxies UGC 128 and Malin 1 over an extended range.\footnote{For galaxies the $N^*\gamma^*$ term in (\ref{M68}) is never larger than of order the $\gamma_0$ term since for galaxies the number of stars $N^*$ is never bigger than of order $10^{11}$. Hence for galaxies the maximum size associated with the distance in which the right-hand side of (\ref{M68}) vanishes is never larger than of order 100 kiloparsec. For clusters of galaxies $N^*$ is of order 1000 times larger as clusters typically contain 1000 galaxies. From (\ref{M68}) the maximum allowed size for clusters is then well into the megaparsec region.}

It is important to appreciate that the fits provided by conformal gravity (and likewise those provided by other  alternate theories such as Milgrom's MOND theory \cite{Milgrom1983} and Moffat's MSTG/SVTG theory \cite{Moffat2005}) are predictions. Specifically, for all these theories the only input one needs is the photometric and HI gas data, and the only free parameter is the $M/L$ ratio for each given galaxy, with rotation velocities then being determined. That these highly constrained alternate theories all work is because not only do they each possess an either derived or postulated underlying universal scale (a derived $\gamma_0=2(-K)^{1/2}=3.06\times 10^{-30}~{\rm cm}^{-1}$ for conformal gravity,  $a_0/c^2=1.33\times 10^{-29}~{\rm cm}^{-1}$ for MOND and $G_0M_0/r_0^2c^2=7.67\times 10^{-29}~{\rm cm}^{-1}$ for MSTG), all of the 138 galaxies in the sample possess it too. Specifically, despite the huge variation in luminosity and  surface brightness across the 138 galaxy sample, within one order of magnitude the measured values of the centripetal accelerations $(v^2/c^2R)_{\rm last}$ at the last data point in each galaxy are all found to cluster around a value of $3\times 10^{-30}~{\rm cm}^{-1}$ or so. For the 21 large galaxy sample for instance the values for  $v^2/c^2R$ all lie within the narrow range $(0.97 - 5.83) \times 10^{-30}~{\rm cm}^{-1}$. 

It should also be noted that while the fits provided by conformal gravity are predictions, in contrast,  dark matter fitting to galactic data works quite differently. There one first needs to know the velocities so that one can then ascertain the needed amount of dark matter, i.e. in its current formulation dark matter is only a parametrization of the velocity discrepancies that are observed and is not a prediction of them. Dark matter theory has yet to develop to the point where it is able to predict rotation velocities given a knowledge of the luminous distribution alone (or explain the near universality found for $(v^2/c^2R)_{\rm last}$). Thus dark matter theories, and in particular those theories that produce dark matter halos in the early universe, are currently unable to make an a priori determination as to which halo is to go with which particular luminous matter distribution, and need to fine-tune halo parameters to luminous parameters galaxy by galaxy. In the NFW CDM simulations \cite{Navarro1996} for instance, one finds generic spherical halo profiles close in form to $\sigma(r)=\sigma_0/[r(r+r_0)^2]$ (as then cut off at some $r_{\rm max}$), but with the halo parameters $\sigma_0$, $r_0$ and $r_{\rm max}$ needing to be fixed galaxy by galaxy. In addition to the galactic mass to light ratios, this requires 414 further parameters for the 138 galaxy sample (or a further 276 parameters for isothermal halo type models). No such fine-tuning shortcomings appear in conformal gravity, and if standard gravity is to be the correct  description of gravity, then a universal formula akin to the one given in (\ref{M67}) and the existence of the universal $\gamma_0$ and $\kappa$ parameters would need to be derived by dark matter theory. 

The conformal gravity fits are also noteworthy in that conformal gravity was not at all developed for the purpose of addressing the dark matter problem. Rather, it was first advocated by the present author \cite{Mannheim1990} solely because it has a symmetry that could address the cosmological constant problem. However, once the starting action of (\ref{M9}) is assumed, one can then proceed purely deductively and derive the rotation curve formula given in (\ref{M67}), a thus purely theoretical first principles approach. Moreover, since our study of (\ref{M67}) then establishes that global physics has an influence on local galactic motions, the invoking of dark matter in galaxies could potentially be nothing more than an attempt to describe global physics effects in purely local galactic terms.

\section{Connecting Conformal and Einstein Gravity}
\label{S10}

In a recent paper 't Hooft \cite{'tHooft2010} has found an interesting connection between Einstein gravity and conformal gravity. In standard treatments of quantum Einstein gravity one makes a perturbative expansion in the metric and generates multi-loop Feynman diagrams. Each perturbative order requires a new counter-term, with the nth-order one being a function of the nth-power of the Riemann tensor and its contractions. With the series not terminating, Einstein gravity is rendered non-renormalizable. 

In his paper 't Hooft proposes a very different approach, one that is highly nonlinear. Specifically, instead of evaluating the path integral as a perturbative series in the metric components $g_{\mu\nu}(x)$, he proposes to treat the conformal factor in the metric as an independent degree of freedom. Specifically, he makes a conformal transformation on the (non-conformal invariant) Einstein-Hilbert action of the form $g_{\mu\nu}(x)= \omega^2(x)\hat{g}_{\mu\nu}(x)$, to obtain 
\begin{equation}
I_{\rm EH}=-{1 \over 16\pi G}\int d^4x (-\hat{g})^{1/2}\left(\omega^2\hat{R}^{\alpha}_{\phantom{\alpha}\alpha} -6\hat{g}^{\mu\nu}\partial_{\mu}\omega\partial_{\nu}\omega\right),
\label{M70}
\end{equation}
with everything in $I_{\rm EH}$ now being evaluated in a geometry with metric $\hat{g}_{\mu\nu}(x)$.\footnote{In deriving (\ref{M70}) we note that if set $g_{\mu\nu}(x)= \omega^2(x)\hat{g}_{\mu\nu}(x)$ we obtain $(-g)^{1/2}R^{\alpha}_{\phantom{\alpha}\alpha}(g_{\mu\nu})=(-\hat{g})^{1/2}[\omega^{2}\hat{R}^{\alpha}_{\phantom{\alpha}\alpha}(\hat{g}_{\mu\nu})+6\bar{g}^{\alpha\beta}\omega\hat{\nabla}_{\alpha}\hat{\nabla}_{\beta}\omega]$ where both $\hat{R}^{\alpha}_{\phantom{\alpha}\alpha}(\hat{g}_{\mu\nu})$ and the $\hat{\nabla}_{\alpha}$ derivatives are evaluated in a geometry with metric $\hat{g}_{\mu\nu}(x)$. An integration by parts then yields (\ref{M70}).}  Then instead of taking the path integral measure to be of the standard $Dg_{\mu\nu}$ form, 't Hooft proposes that it be taken to be of the form $D\omega D(g_{\mu\nu}/\omega^{2})=D\omega D\hat{g}_{\mu\nu}$. 

The utility of this approach is that since the $\omega$ dependence in $I_{\rm EH}$ is quadratic, the $D\omega$ path integral can be done analytically. However, in order for the path integral to be bounded one needs to give $\omega$ an imaginary part.  With this choice, the $\omega$ path integral will generate an effective action $I_{\rm EFF}$ of the form ${\rm Tr~ln}[\hat{g}^{\mu\nu}\hat{\nabla}_{\mu}\hat{\nabla}_{\nu}+(1/6)\hat{R}^{\alpha}_{\phantom{\alpha}\alpha}]$, and after dimensional regularization  is found not to generate an infinite set of divergent terms at all, but rather to only generate just one divergent term, viz. the logarithmically divergent
\begin{equation}
I_{\rm EFF}=\frac{C}{120}\int d^4x (-\hat{g})^{1/2}[\hat{R}^{\mu\nu}\hat{R}_{\mu\nu}-{1 \over 3}(\hat{R}^{\alpha}_{\phantom{\alpha}\alpha})^2],
\label{M71}
\end{equation}
with $C$ being the very same logarithmically divergent constant that had appeared in (\ref{M16}).

In (\ref{M71}) we immediately recognize the conformal gravity action. Since the action in (\ref{M70}) is the same action as that obeyed by a conformally coupled scalar field, and the $\omega$ path integral measure is the same as that of a scalar field, everything is conformal, and the $\omega$ path integral can only generate a conformal invariant effective action -- hence (\ref{M71}).  Through the unusual treatment of the conformal factor we thus find a connection between Einstein gravity and conformal gravity. 

From the perspective of Einstein gravity, the utility of (\ref{M71}) is that since conformal gravity is renormalizable, the subsequent $D\hat{g}_{\mu\nu}$ integration of $I_{\rm EFF}$ should not generate any additional counter-terms, while the non-leading terms contained in ${\rm Tr~ln}[\hat{g}^{\mu\nu}\hat{\nabla}_{\mu}\hat{\nabla}_{\nu}+(1/6)\hat{R}^{\alpha}_{\phantom{\alpha}\alpha}]$ would generate contributions to the path integration that could potentially be finite. One still has to deal with the divergent $C$ term in (\ref{M71}), and rather than have it renormalized (say by adding on an intrinsic conformal $I_{\rm W}$ term), 't Hooft explores the possibility that $C$ remain uncanceled, and we refer the reader to his paper for details. (With $C$ appearing with the same overall sign in (\ref{M16}) and (\ref{M71}), and with a gauge field path integration over its kinetic energy having the same sign too \cite{'tHooft2010}, an interplay between fermionic and bosonic fields cannot cancel $C$ -- with massless superpartner fields not being able to effect the same cancellation in a curved background that they can achieve in a flat one.)

While we thus generate a conformal action if we start with an Einstein action, from the perspective of pure conformal gravity, conformal invariance prevents one from including an Einstein term in the fundamental action at all.\footnote{While conformal invariance excludes a fundamental Einstein-Hilbert action, various authors have  discussed the possibility that one might be induced or derived starting from the conformal gravity action. See e.g. S.~L.~Adler,~\emph{ Rev.~Mod.~Phys.}~\textbf{ 54}, 729 (1982); A.~Zee,~\emph{Ann.~Phys.~(N.~Y.)}~\textbf{151}, 431 (1983); J.~Maldacena, {\it Einstein gravity from conformal gravity}, arXiv:1105.5632 [hep-th].} However, in a pure conformal theory one still needs to use a measure that is conformal invariant. Now the metric itself is not conformal invariant, and hence neither is the $Dg_{\mu\nu}$ path integration measure. However, the quantity $g_{\mu\nu}/(-g)^{1/4}$ is conformal invariant and thus so is an integration measure of the form $D(g_{\mu\nu}/(-g)^{1/4})$ (and analogously $D((-g)^{3/16}\bar{\psi})D((-g)^{3/16}\psi)$ for fermions). In order to simplify the measure  one would like to work in a conformal gauge in which the determinant of the metric is fixed to a convenient value such as one. And while it needs to be explored in detail, it is possible that adding an Einstein term to the conformal action and taking the measure to be of the form $D(-g)^{1/8}D(g_{\mu\nu}/(-g)^{1/4})$ might then serve as an appropriate conformal gauge fixing procedure. The fact that there would only be nine independent $D(g_{\mu\nu}/(-g)^{1/4})$ terms parallels the fact that the perturbative $I_{\rm W}$ of (\ref{M22}) only depends on the traceless 9-component $K^{\mu\nu}=h^{\mu\nu}-(1/4)\eta^{\mu\nu}\eta_{\alpha\beta}h^{\alpha\beta}$. Additionally, the fact that one needs to give $\omega$ an imaginary part in order to obtain a well-defined path integral parallels the structure we found  for the conformal theory, with unitarity being achieved by having the gravitational field be anti-Hermitian.

\section{Summary}
\label{S11}

The present paper is the fifth in a series of publications in the journal on conformal gravity by the author, with much of the earlier work on the conformal theory having been reviewed in the previous four \cite{Mannheim1994,Mannheim1996,Mannheim2000,Mannheim2007}. In this review we have discussed some recent developments in the field and presented the case that can currently be made for conformal gravity. In particular we have shown how the conformal theory can quite naturally handle some of the most troublesome problems in physics, the quantum gravity problem, the vacuum energy problem, and the dark matter problem, being able to do so in the four spacetime dimensions for which there is evidence. As detailed in \cite{Mannheim2006} much more still needs to be done: anisotropy of  the cosmic microwave background, large scale structure, cluster dynamics and lensing by clusters (especially in light of the recently found global $-\kappa c^2 R$ term in (\ref{M67}) and (\ref{M68})), orbit decays of binary pulsars and gravity waves, solving the primordial deuterium problem that conformal nucleosynthesis has.\footnote{Many of these issues involve the growth of cosmological inhomogeneities and their interplay with the cosmological background as exhibited in (\ref{M67}). A first step toward addressing these issues and in developing conformal cosmological perturbation theory in general has recently been taken in P.~D.~Mannheim,~{\it Cosmological perturbations in conformal gravity}, arXiv:1109.4119 [gr-qc]. It will be of interest to ascertain the degree to which conformal cosmological fluctuation theory is aware of the $\gamma_0$ and $\kappa$ scales, especially since $\kappa$ is a matter fluctuation moment integral.} For all of these applications we only need to consider the particle contribution to the finite (\ref{M32}), with the contribution of the vacuum sector including the cosmological constant having been taken care of by (\ref{M31}). All of these issues should eventually prove definitive for the conformal theory, especially since it has none of the freedom associated with the difficult to pin down and still highly elusive dark matter and dark energy present in the standard theory.\footnote{In passing we note that even if some dark matter candidate particles are found in an accelerator experiment, to establish that such particles contribute to a possible dark matter halo in the Milky Way Galaxy, one would need to determine their local galactic density. For dark matter to not be excluded, the parameter space allowed by dark matter detection in an accelerator experiment would have to not conflict with the parameter space that has already been excluded by the non-detection to date of dark matter in non-accelerator experiments.} The highly constrained conformal gravity fits to galactic rotation curves have as yet no parallel in dark matter theory where parameters need to be fine-tuned galaxy by galaxy, and its natural solution to the cosmological constant problem has as yet no parallel in standard cosmology where $\Lambda$ needs to be fine-tuned to a degree without precedent in physics. To conclude, we note that at the beginning of the 20th century studies of black-body radiation on microscopic scales led to a paradigm shift in physics. Thus it could that at the beginning of the 21st century studies of black-body radiation, this time on macroscopic cosmological scales, might be presaging a paradigm shift all over again.

\begin{acknowledgements}

This paper is based in part on a presentation made by the author at the International Conference on Two Cosmological Models, Universidad Iberoamericana, Mexico City, November 2010. The author wishes to thank Dr. J. Auping and Dr. A. V. Sandoval for the kind hospitality of the conference.

\end{acknowledgements}

\vfill\eject

\begin{figure}
\epsfig{file=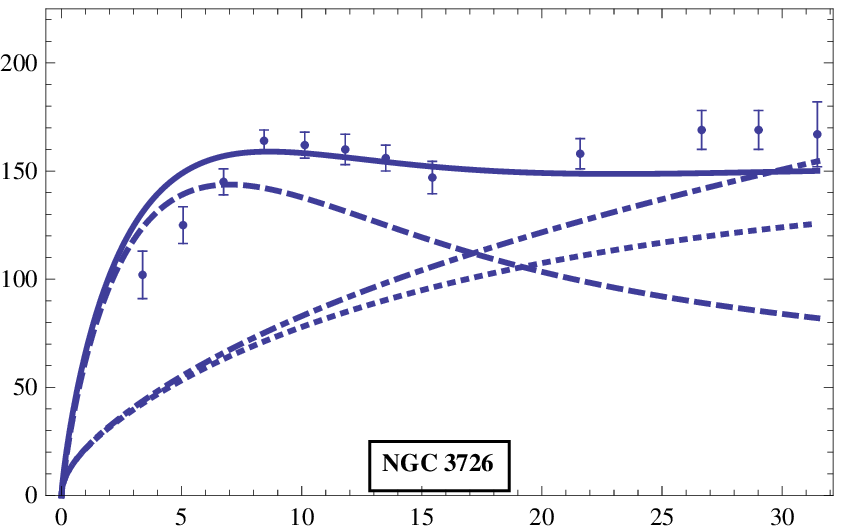,width=1.42in,height=0.85in}~~~
\epsfig{file=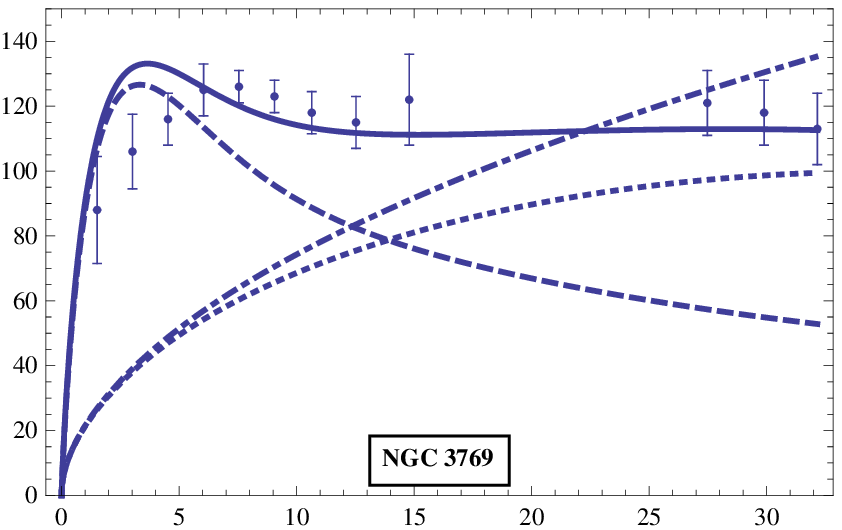,width=1.42in,height=0.85in}~~~
\epsfig{file=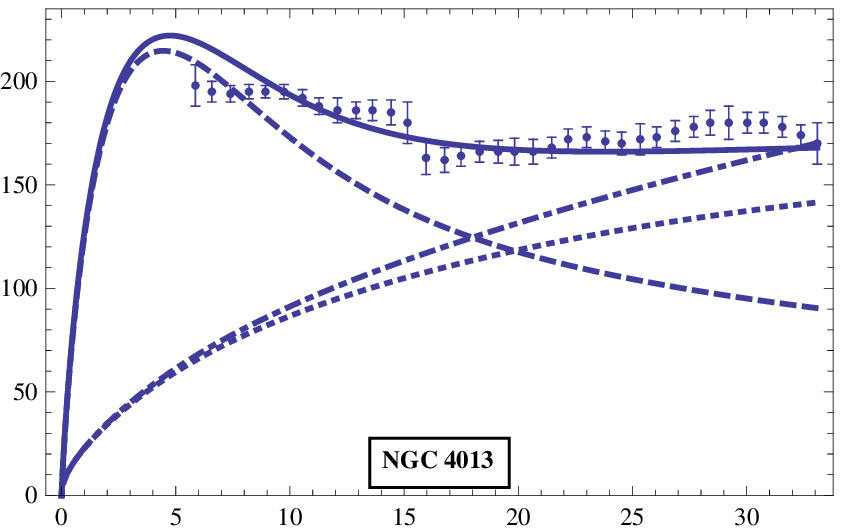,width=1.42in,height=0.85in}
\end{figure}

\begin{figure}
\epsfig{file=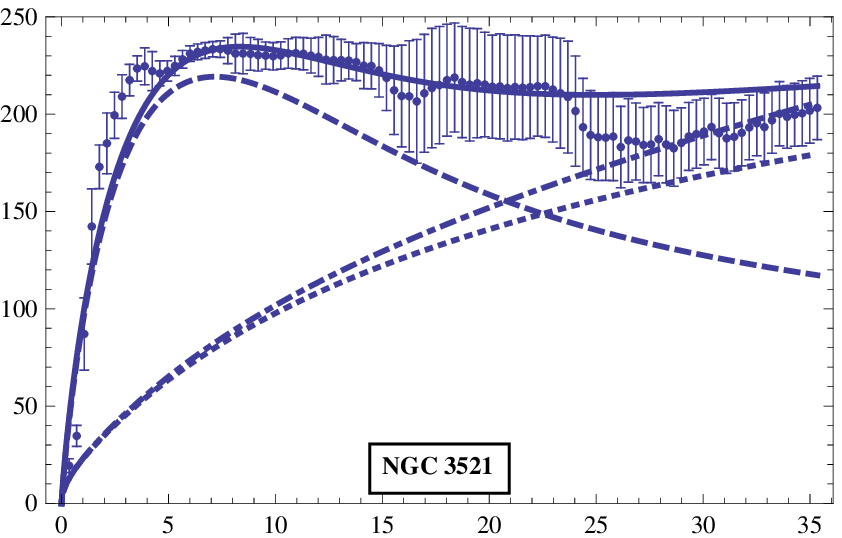,width=1.42in,height=0.85in}~~~
\epsfig{file=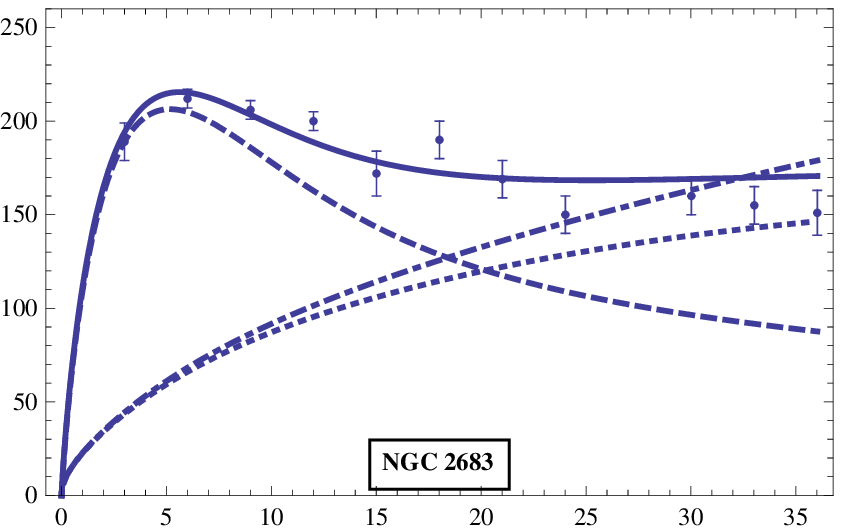,width=1.42in,height=0.85in}~~~
\epsfig{file=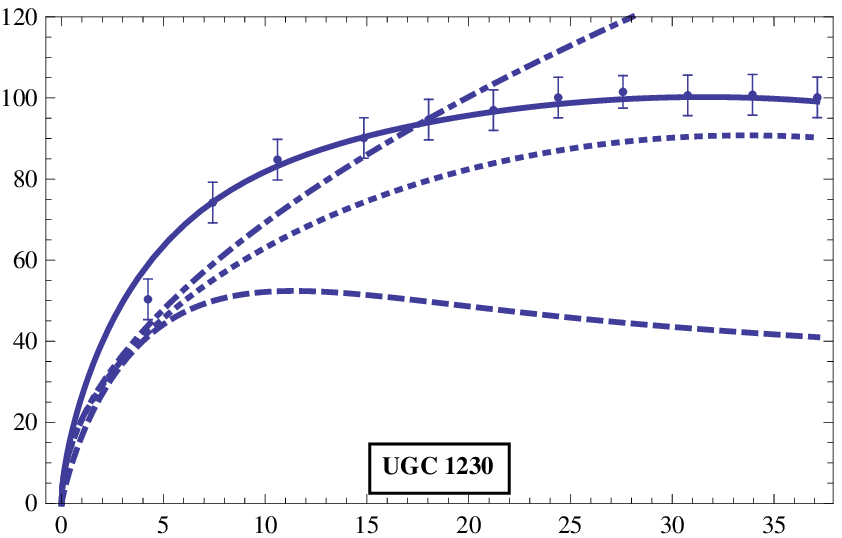,width=1.42in,height=0.85in}
\end{figure}

\begin{figure}
\epsfig{file=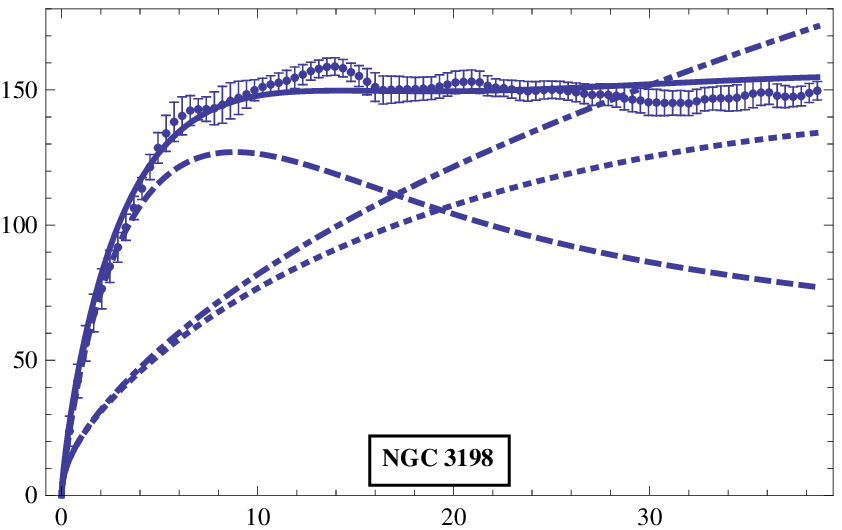,width=1.42in,height=0.85in}~~~
\epsfig{file=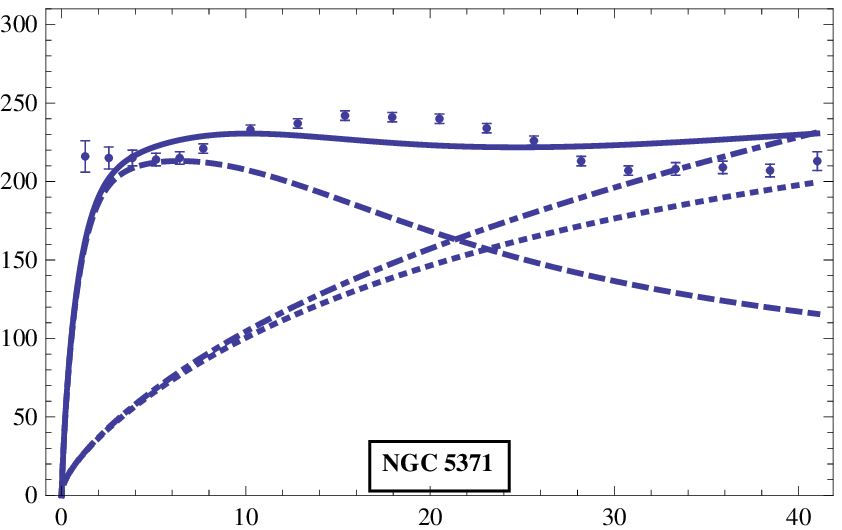,width=1.42in,height=0.85in}~~~
\epsfig{file=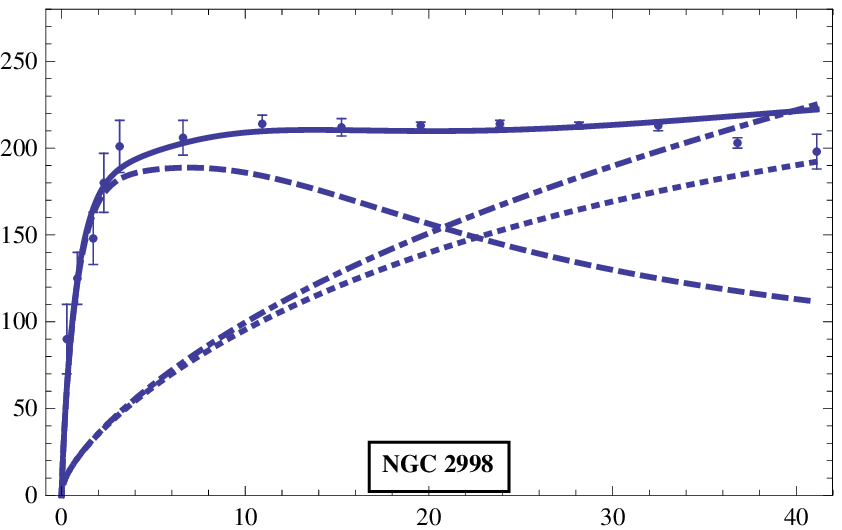,width=1.42in,height=0.85in}
\end{figure}

\begin{figure}
\epsfig{file=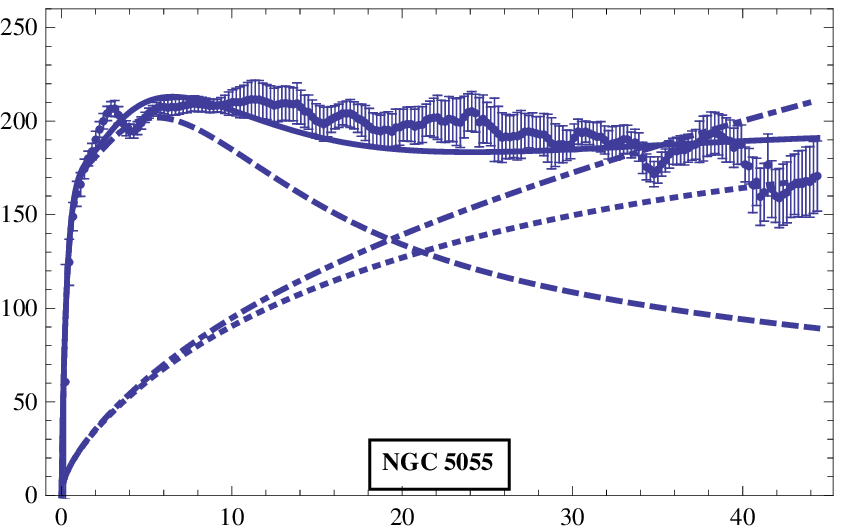,width=1.42in,height=0.85in}~~~
\epsfig{file=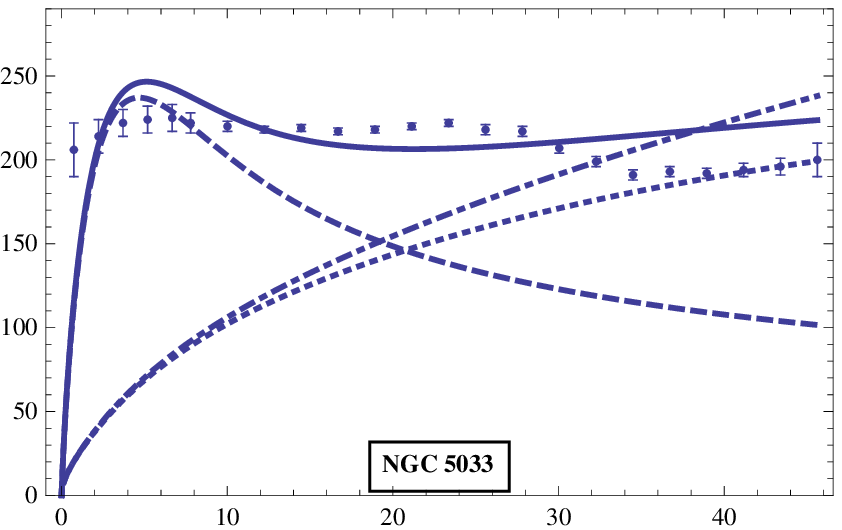,width=1.42in,height=0.85in}~~~
\epsfig{file=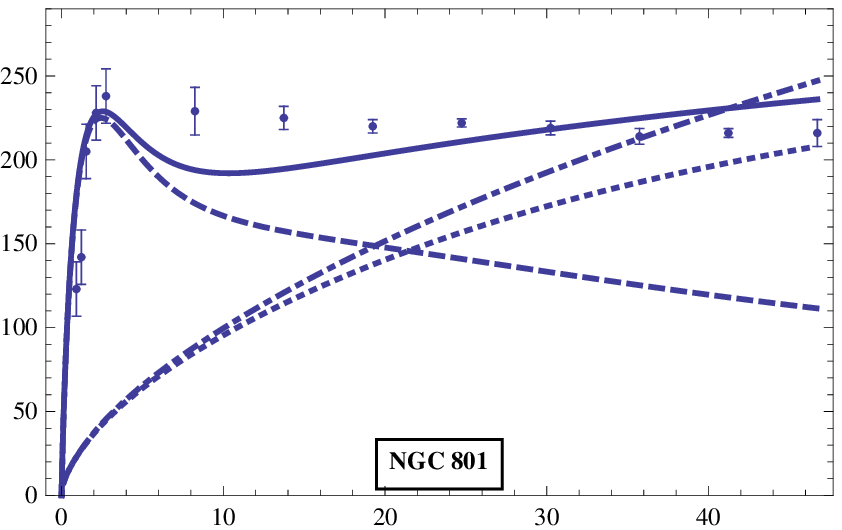,  width=1.42in,height=0.85in}
\end{figure}

\begin{figure}
\epsfig{file=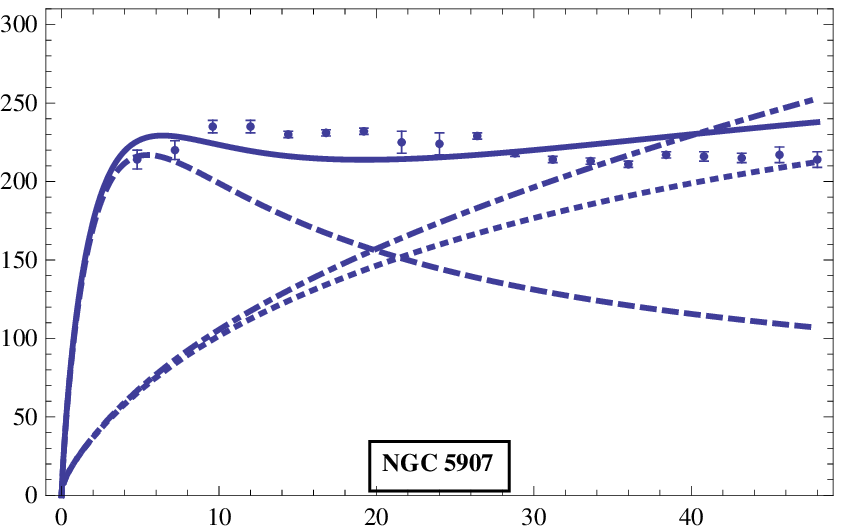,width=1.42in,height=0.85in}~~~
\epsfig{file=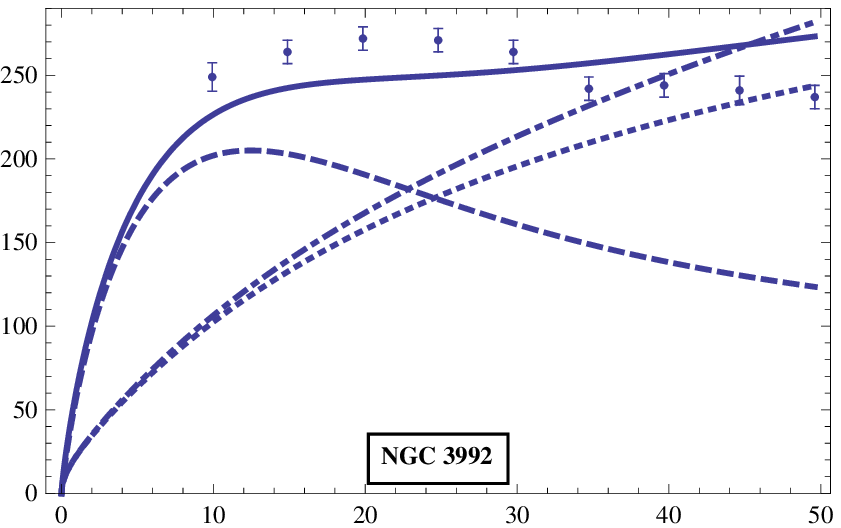,width=1.42in,height=0.85in}~~~
\epsfig{file=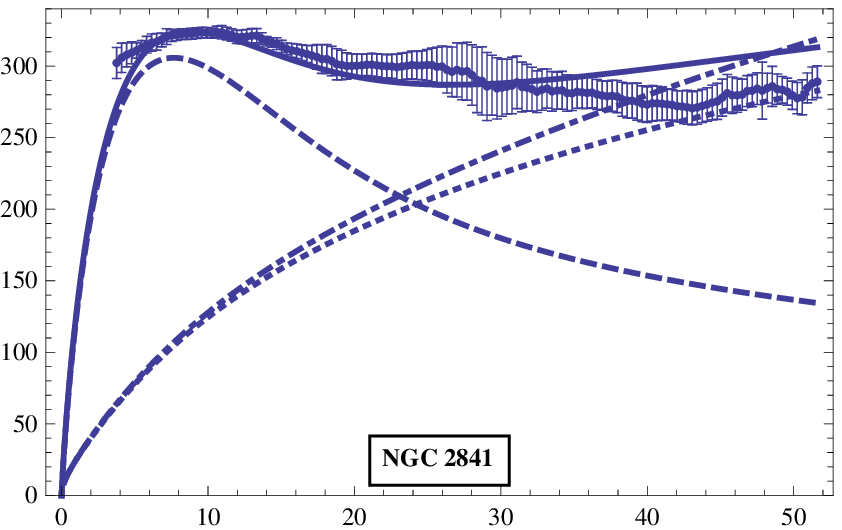,width=1.42in,height=0.85in}
\end{figure}

\begin{figure}
\epsfig{file=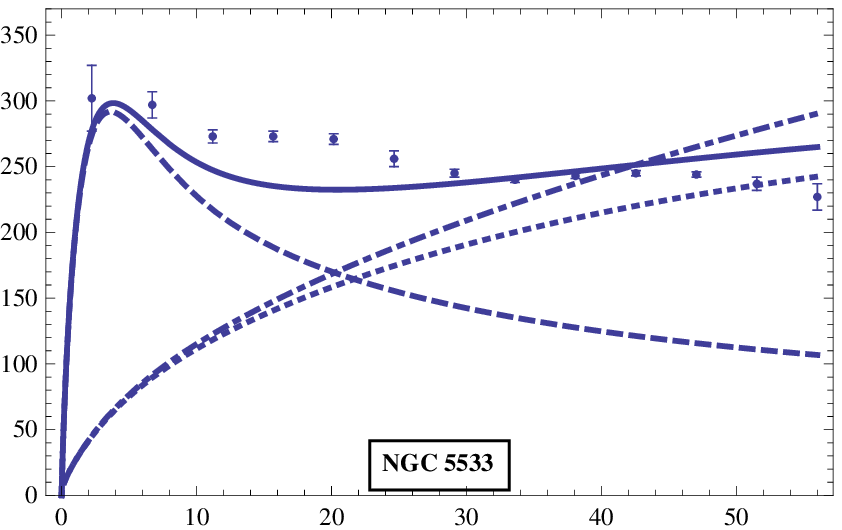,width=1.42in,height=0.85in}~~~
\epsfig{file=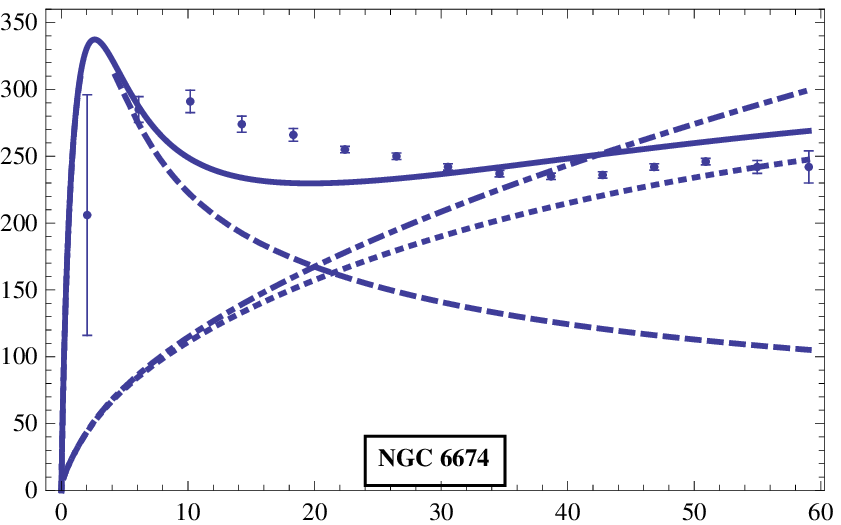,width=1.42in,height=0.85in}~~~
\epsfig{file=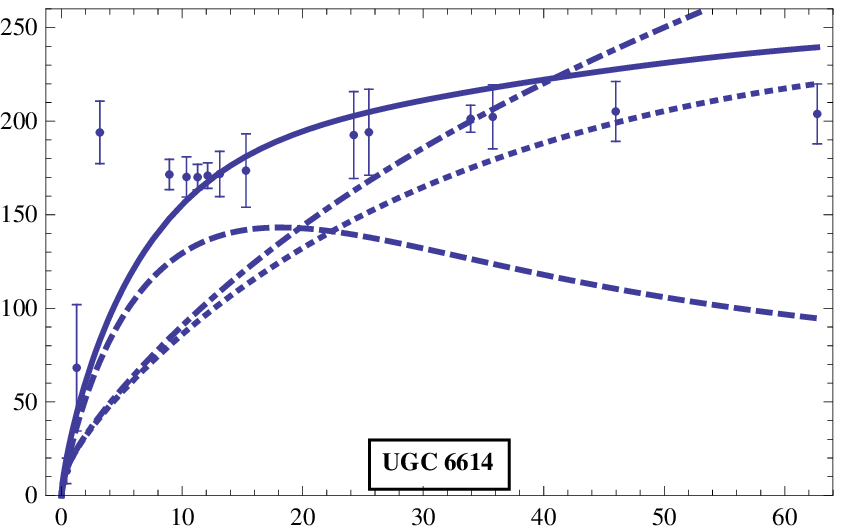,width=1.42in,height=0.85in}
\end{figure}

\begin{figure}
\epsfig{file=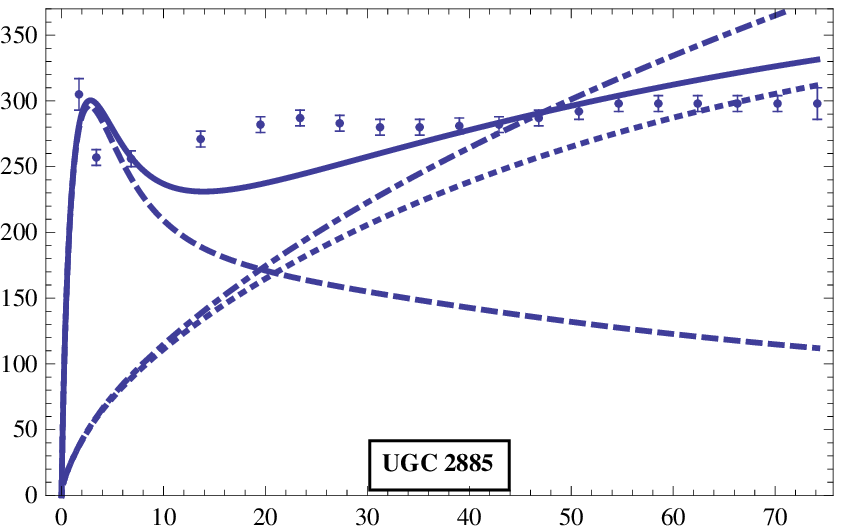,width=1.42in,height=0.85in}~~~
\epsfig{file=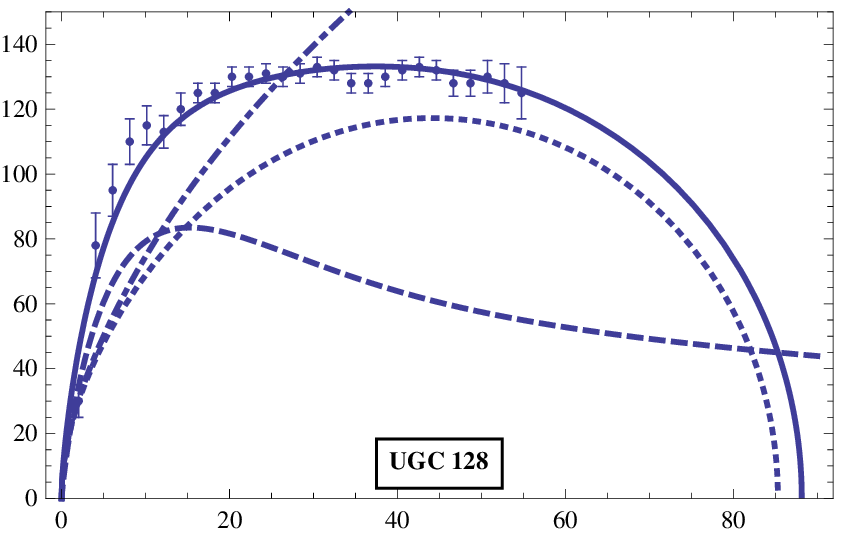, width=1.42in,height=0.85in}~~~
\epsfig{file=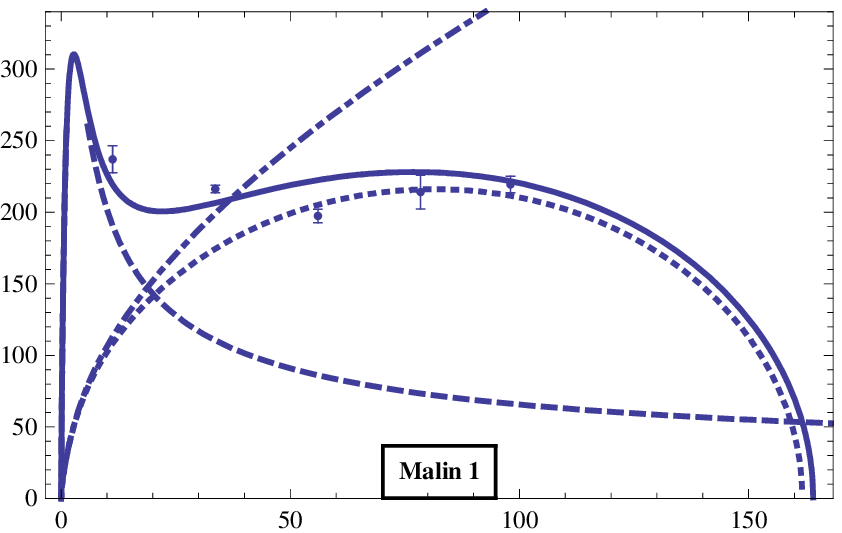, width=1.42in,height=0.85in}\\\
\medskip
\caption{Fitting to the rotational velocities in km~sec$^{-1}$ as a function of radial distance in kpc}
\label{Fig. (1)}
\end{figure}

\end{document}